\documentclass[11pt,a4paper,leqno]{article}
\usepackage[utf8]{inputenc}
\usepackage{authblk}
\usepackage{geometry}
\usepackage{natbib}
\usepackage{graphicx}
\usepackage{pdflscape}
\usepackage{appendix}
\usepackage{threeparttable}
\usepackage{booktabs}
\usepackage{amssymb}
\usepackage{amsbsy}
\usepackage{bm}
\usepackage{amsmath}
\usepackage{amsthm}
\usepackage{graphicx}
\usepackage{mathtools}
\usepackage{rotating}
\usepackage{setspace}
\usepackage{diagbox}
\usepackage{footmisc}

\usepackage[table]{xcolor}    

\usepackage{colortbl}
\providecommand{\shadeRow}{\rowcolor[rgb]{0.9, 0.9, 0.9}}
\providecommand{\shadeBench}{\rowcolor[RGB]{255, 195, 180}}

\usepackage{tikz}
\usepackage{enumitem}
\newlist{steps}{enumerate}{1}
\setlist[steps, 1]{label = Step \arabic*:}

\usepackage{dcolumn}
\newcolumntype{d}[1]{D{.}{.}{#1}}

\usepackage{caption}
\usepackage{subcaption}
\definecolor{nblue}{HTML}{000660}
\usepackage[colorlinks=true,urlcolor=nblue,linkcolor=nblue,citecolor=nblue]{hyperref}

\definecolor{Florian}{RGB}{255,0,0}

\title{\textbf{Approximate Bayesian inference and forecasting in huge-dimensional multi-country VARs}\footnote{This paper is set to appear in \textit{International Economic Review} subject to editorial changes.}}

\author{}
\date{}

\begin{document}

\maketitle
\thispagestyle{empty}
\vspace*{-6.5em} 
\begin{center}
\end{center}

\normalsize

\vspace*{1em}
\begin{minipage}{.49\textwidth}
  \large\centering Martin Feldkircher\\[0.25em]
  \small Vienna School of International Studies, Austria
\end{minipage}
\begin{minipage}{.49\textwidth}
  \large\centering Florian Huber\footnotemark{}\\[0.25em]
  \small University of Salzburg, Austria
\end{minipage}

\vspace*{1.5em}

\begin{minipage}{.49\textwidth}
  \large\centering Gary Koop\\[0.25em]
  \small University of Strathclyde, United Kingdom
\end{minipage}
\begin{minipage}{.49\textwidth}
  \large\centering Michael Pfarrhofer\\[0.25em]
  \small University of Salzburg, Austria
\end{minipage}
\footnotetext{
We thank the editor Jesús Fernández-Villaverde and two anonymous referees for their valuable suggestions and constructive  comments. 
We also thank Maximilian B\"ock, Todd Clark, Niko Hauzenberger, Ed Knotek, James Mitchell, Anna Stelzer, Saeed Zaman, and participants of the IAAE 2021 Annual Conference in Rotterdam. Florian Huber and Michael Pfarrhofer gratefully acknowledge financial support from the Austrian Science Fund (FWF, grant no. ZK 35) and the Jubiläumsfonds of the Oesterreichische Nationalbank (grant no. 18304). Please address correspondence to: Florian Huber. Department of Economics, University of Salzburg. \textit{Address}: M\"{o}nchsberg 2a, 5020 Salzburg, Austria. \textit{Email}: \href{mailto:florian.huber@plus.ac.at}{florian.huber@plus.ac.at}.}
\vspace*{2em}
\begin{center}
\begin{minipage}{0.925\textwidth}
\begin{abstract}

\setcounter{page}{0}
\doublespacing
\noindent Panel Vector Autoregressions (PVARs) are a popular tool for analyzing multi-country datasets. However, the number of estimated parameters can be enormous, leading to computational and statistical issues. In this paper, we develop fast Bayesian methods for estimating PVARs using integrated rotated Gaussian approximations. We exploit the fact that domestic information is often more important than international information and group the coefficients accordingly. Fast approximations are used to estimate the latter while the former are estimated with precision using Markov chain Monte Carlo techniques. We illustrate, using a huge model of the world economy, that it produces competitive forecasts quickly.
\end{abstract}
\end{minipage}
\end{center}
\begin{center}
\begin{minipage}{0.825\textwidth}
\bigskip
\begin{tabular}{p{0.2\hsize}p{0.65\hsize}} 
\textbf{Keywords:} &  Multi-country models, macroeconomic forecasting, vector autoregression, spillovers.\\
\textbf{JEL Codes:} & C11, C33, C55, E37 \\
\end{tabular}
\end{minipage}
\end{center}
\normalsize

\newpage\doublespacing

\section{Introduction}
There is much evidence that working with multi-country time series models improves macroeconomic forecasting and structural analysis \citep[see, among many others,][]{psw2004,pss2009,cc2016}.
This is to be expected in the modern globalized economy where countries are linked together through trade and financial flows and events in one country can spill over into others. However, the relevant data sets can be enormous. In this paper, we work with a $38$ country data set that contains $487$ variables. If all of these are treated as endogenous variables in an unrestricted multi-country vector autoregression (VAR), the number of equations in the VAR will be huge, as will the number of right hand side variables in each equation. The resulting model will be over-parameterized.  

Much of the existing literature deals with this problem by imposing restrictions on the model or compressing the data (e.g., by using factor methods). For instance, the popular class of global VARs \citep[GVARs, see, e.g.,][]{psw2004,pss2009,cfh2016,dfh2016,huber2016} assumes that information from all other countries impacts a country solely through a single weighted average of other country information. The weights in the average typically are based on bilateral trade flows. By contrast, the literature on panel VARs (PVARs) mainly deals with over-parameterization issues through constraints on the parameters describing the dynamic and static relations across countries \citep[see, e.g.,][]{canova2009estimating}. 
 
We propose an unrestricted PVAR specification where any variable can affect any or all other variables either contemporaneously or with a lag. This implies that the influence of foreign variables is neither a priori restricted (e.g., using trade weights) nor based on the output of a dimension-reduction procedure such as principal components. In other words, we want to let the data decide the exact nature and extent of linkages between countries. This feature of our approach turns out to be a substantial improvement over multi-country models commonly used in the literature.
 
Bayesian shrinkage methods are increasingly used for overcoming the over-parameterization problems which arise when working with unrestricted PVARs (or large VARs in general). Influential early contributions such as \cite{canova2009estimating} used simple methods for choosing the prior (e.g., subjective elicitation or training sample methods). More recently, Bayesians have been working with global-local shrinkage or variable selection priors \citep[see, e.g.,][]{kk2016, korobilis2016prior, bai2022}. These priors are commonly used in Big Data problems where models involve a large number of parameters. They automatically sort through all the parameters and decide which ones to shrink to zero and which ones to estimate freely. Our goal in this paper is to use a global-local shrinkage prior in a large unrestricted PVAR and let it decide which cross-country linkages are important and which can be ignored. In our empirical work, we mainly use the Horseshoe prior of \cite{horseshoe} although the econometric methods developed in this paper will work with any hierarchical shrinkage prior that takes a conditionally Gaussian form including the LASSO and Dirichlet-Laplace priors of \cite{lasso} and \cite{bhattacharya2015dirichlet}, respectively.

Bayesian estimation and forecasting in VARs using global-local shrinkage priors typically requires the use of computationally demanding Markov Chain Monte Carlo (MCMC) methods. These algorithms are impractical in the very large PVARs that arise when working with many countries and many variables. For this reason, the existing Bayesian literature which uses unrestricted PVARs has focused on relatively small models. For instance, \cite{kk2016} use a PVAR involving four variables for each of seven European countries which is much smaller than the one considered in this paper. 
    
To overcome the computational hurdle, we develop an Integrated Rotated Gaussian Approximation (IRGA) for the PVAR. IRGAs were recently developed by \cite{irga} as a machine learning tool to speed up computation in high-dimensional models. These methods build on the intuition that some parameters are more important than other parameters. The other parameters, which in our case are mainly coefficients associated with other countries' lagged endogenous variables and covariance terms, are estimated using efficient approximations. The more important parameters are then estimated conditional on these approximations using precise MCMC techniques. We adapt these methods for use with PVARs. The resulting IRGA-based algorithm leads to vast improvements in speed of computation and has appealing approximation properties which we illustrate through simulations.

In our empirical work, we estimate a huge model of the world economy that contains 487 endogenous variables. In a forecasting exercise,  we compare the performance of our unrestricted PVAR with a Horseshoe prior, estimated using IRGA methods (PVAR-IRGA), to a range of alternatives including a GVAR, factor-augmented VARs with the factors constructed using other country variables, and single country Bayesian VARs. In terms of computation, our key finding is that the computational improvements are large enough to enable Bayesian forecasting and structural analysis to be done even in huge PVARs.  In terms of empirical results, we find our large approximate model forecasts  well and often outperforms competing models. The forecast improvements are particularly strong for short-run density forecasts of stock market returns and longer-run inflation and output predictions. In these cases, we find  PVAR-IRGA, with few exceptions, to forecast substantially better than all the alternatives. 

We then proceed by analyzing the properties of the forecasts of the PVAR-IRGA via recent techniques used for analyzing social networks \citep{holland1983stochastic, karrer2011stochastic, vziberna2014blockmodeling, pati2015optimal}. These stochastic block models build on the correlation matrix and sort correlations into clusters which facilitate interpretation. This analysis  provides novel insights on the properties of the forecasts which are consistent with actual developments during the global financial crisis and the euro area sovereign debt crisis. 

Our forecasting exercise is complemented by additional empirical results on the degree of cross-country spillovers. Considering different variants of the \citet{dy2009}  spillover index shows that our model detects sizable international relations across output, prices, long-term interest rates and stock markets which sharply increase throughout the hold-out period. These increases are particularly pronounced during times of economic turmoil. Our findings hence provide novel insights on truly global connectivity since comparable large scale analyses are not feasible with standard econometric techniques.

The remainder of the paper is structured as follows. In Section \ref{sec:PVAR} we define the PVAR likelihood and the prior we use to carry out Bayesian inference and prediction. Section \ref{sec:approxIRGA} discusses computation and develops our IRGA methods which allow for fast computation. It also includes a theoretical discussion on the approximation properties of IRGA.  In Section \ref{sec: synth} we carry out a simulation exercise to complement the theoretical discussion while in Section \ref{sec: emp_work}  we present results for our forecasting exercise which compares our PVAR-IRGA to a range of other approaches. This section also includes information on the extent of cross-country spillovers. The final section summarizes and concludes the paper. The appendix provides additional technical details and further empirical results such as robustness checks.

\section{The panel VAR}\label{sec:PVAR}
This section develops the basic PVAR and briefly discusses the main specification issues commonly faced by researchers. We then discuss how Bayesian techniques can be used to deal with over-parameterization concerns and thus, in an automatic fashion, solve several of these specification issues.
\subsection{The likelihood function}
Our goal is to model dynamic and static relations in an international panel of macroeconomic and financial time series which are stored in an $n$-dimensional vector $\bm y_t = (\bm y'_{1t}, \dots, \bm y'_{Nt})'$ for $t=1,\hdots,T$. This  vector is composed of $N$ country-specific  sub-vectors $\bm y_{it}$ which are $M \times 1$ dimensional.\footnote{Note that $M$ may differ across countries but is used here to simplify notation. Our approach naturally allows for different covariates across equations and countries. In the empirical work, variable coverage differs across countries.}  Assuming that each $\bm y_{it}$ depends on the lagged values of $\bm y_t$, we obtain a PVAR  given by:
\begin{equation}
    \bm y_{it} = \bm \Gamma_{i1}\bm y_{it-1} + \dots + \bm \Gamma_{ip}\bm y_{it-p} +  \bm \Xi_i \bm z_{it} + \bm \epsilon_{it},\label{eq:likelihood}
\end{equation}
where $\bm \Gamma_{ij}$ are $M \times M$ coefficient matrices associated with the lagged endogenous variables of country $i$. Lags of variables from countries other than $i$ are denoted by $\bm z_{it} = (\bm y'_{-i, t-1}, \dots, \bm y'_{-i, t-p})'$ with $\bm y_{-i, t}=(\bm y'_{1t}, \dots, \bm y'_{i-1, t}, \bm y'_{i+1, t}, \dots, \bm y'_{Nt})'$. The coefficient matrix on other country lags, $\bm \Xi_i$, is an $M \times K_{other}$ matrix where $K_{other} = (N-1)Mp$. Note that $\bm \Xi_i$ will contain an enormous number of parameters unless $N$ and/or $M$ are small. The matrix $\bm \Xi_i$ encodes the dynamic relations across countries (which are commonly referred to as dynamic interdependencies in the literature) while the  $M \times k(=Mp)$ matrix $\bm \Gamma_i = (\bm \Gamma_{i1}, \dots, \bm \Gamma_{ip})$ captures domestic dynamics. 

The usual VAR representation in terms of $\bm y_t$ is obtained by stacking all country-specific models and reshuffling the columns of $\bm \Gamma_i$ and $\bm \Xi_i$ appropriately:
\begin{equation}
    \bm y_t =  \tilde{\bm \Gamma}_1 \bm y_{t-1} + \dots +  \tilde{\bm \Gamma}_p \bm y_{t-p} + \bm \epsilon_t. \label{eq: global_mod}
\end{equation}
The coefficient matrices $\tilde{\bm \Gamma}_j$ are of dimension $n \times n$ and the errors $\bm \epsilon_t = (\bm \epsilon'_{1t}, \dots, \bm \epsilon'_{Nt})'$ are i.i.d. Gaussian with $\bm \Sigma$ being an $n \times n$-dimensional  variance-covariance matrix. The off-diagonal elements of this matrix determine both contemporaneous relations across variables \textit{within} a country and instantaneous dependencies \textit{across} countries. The latter relations are typically referred to as static interdependencies in the literature. Notice that unrestricted estimation of all contemporaneous relations across countries implies estimating $n (n-1) /2$ covariances. For large $n$, this adds to the already huge number of parameters in the $\bm \Gamma_{i}$'s and $\bm \Xi_i$'s.

In the literature on PVARs, estimation is often facilitated by introducing restrictions \citep[see e.g.,][]{canova2009estimating, cc2016} on the coefficients in (\ref{eq:likelihood}) and (\ref{eq: global_mod}). For instance, the so-called cross-sectional homogeneity restriction arises if $\bm \Gamma_{i} = \bm \Gamma_{s}$ for $i \neq s$. This implies that domestic dynamics across countries are identical -- a rather restrictive assumption if the panel of countries includes, e.g., developed and developing economies. Another restriction often introduced is $\bm \Xi_i = \bm 0$ for some (or even all) $i$. This rules out dynamic relations across some countries but substantially reduces the number of free parameters. GVARs are also not restriction free. The assumption here is that cross-country linkages can be approximated by cross-country weighted averages and hence restrict $\bm \Xi_i$.\footnote{These weights have to be  specified exogenously and are often based on measures of economic connectivity such as bilateral trade flows. For an overview, see \citet{Feldkircher2016a}.}  Another restriction sometimes considered assumes that shocks across countries are uncorrelated, i.e., $Cov(\bm \epsilon_{it}, \bm \epsilon_{st}) = \bm 0$. This implies introducing zero restrictions on the relevant elements in $\bm \Sigma$.  All these restrictions, however, have serious implications for forecasting and structural inference and potentially introduce mis-specification if chosen wrongly. These considerations inspire us to use Bayesian variable selection methods via a global-local shrinkage prior so as to choose the appropriate restrictions in a data based manner. 

If left unrestricted, estimation of the PVAR using traditional Bayesian MCMC methods is computationally  cumbersome. For large data sets such as the one used in this paper, the computational burden becomes impractical. Hence, a goal of this paper is to speed up computation. In a first step, we greatly simplify computation  by transforming the PVAR to allow for equation-by-equation estimation. This can be achieved by taking a Cholesky-type decomposition of $\bm \Sigma = \bm U \bm H \bm U'$. Here, we let $\bm U$ denote a lower triangular matrix with unit diagonal and $\bm H$ is a diagonal matrix with main diagonal $\bm \sigma^2 = (\bm \sigma_1^{2'}, \dots, \bm \sigma_N^{2'})'$. The $M$-dimensional vector $\bm \sigma_i^{2} = (\sigma_{\varepsilon, i1}^2, \dots, \sigma_{\varepsilon, iM}^2)'$ stores the idiosyncratic variances $\sigma_{\varepsilon, ij}^2$ associated with the shock in country $i$ and equation $j$.  We can use this decomposition to recover the structural form of (\ref{eq: global_mod}):
\begin{equation*}
    \bm y_t = \bm A_1 \bm y_{t-1} + \dots + \bm A_p \bm y_{t-p} + \bm W \bm y_t + \bm \varepsilon_t, \quad \bm \varepsilon_t \sim \mathcal{N}(\bm 0, \bm H),
\end{equation*}
where $\bm W = (\bm I - \bm U^{-1})$ encodes the contemporaneous relations across the shocks in the system and the matrices $\bm A_j~(j=1,\dots, p)$ denote structural coefficients. Within a given country, we can easily obtain a representation similar to (\ref{eq:likelihood}) by reshuffling the explanatory variables:
 \begin{equation*}
     \bm y_{it} = \bm A_{i1}\bm y_{it-1} + \dots + \bm A_{ip}\bm y_{it-p} +  \bm B_i \bm z_{it} + \bm W_i \bm y_t + \bm \varepsilon_{it},\label{eq:struc_likelihood}
 \end{equation*}
 with $\bm W_i$ denoting the $M$ rows of $\bm W$ associated with the $i^{th}$ country. This  allows us to rewrite the $j^{th} (>1)$ equation in the country-specific model $i (>1)$ as follows:\footnote{For $j=1$ and $i=1$, the equation simplifies and only $\bm x_{it}$ and $\bm z_{it}$ appear as regressors.}
        \begin{equation}
            y_{ij, t} = \bm A'_{ij, \bullet} \bm x_{it} + \bm B'_{ij, \bullet} \bm z_{it}  + \sum_{s=1}^{j-1} w_{is} y_{is, t} + \sum_{v = 1}^{i-1} \bm u'_{iv} \bm y_{vt} +   \varepsilon_{ij, t} \label{eq: irga_eq}
        \end{equation}
where $y_{ij, t}$ denotes the $j^{th}$ element of $\bm y_{it}$, $\bm x_{it} = (\bm y'_{it-1}, \dots, \bm y'_{it-p})'$ while $\bm A_{ij, \bullet}$ and $\bm B_{ij, \bullet}$ denote the $j^{th}$ rows of $\bm A_i$ and $\bm B_i$, respectively.    $\bm u_{ij, \bullet} = (w_{i1}, \dots, w_{i, j-1}, \bm u'_{i1}, \dots, \bm u'_{i i-1})'$  are the covariance parameters associated with the relevant row in $\bm W_i$. 

Note that the errors are now independent across equations (i.e., both within and across countries) with error variance given by $\sigma^2_{\varepsilon, ij}$. This independence allows for estimating one equation at a time which greatly speeds up computation. Cholesky-based formulations such as this have been used in many recent papers, including \cite{ccm2019}, \cite{kkp2019}, \cite{hko2020} and \citet{carriero2021corrigendum}. As opposed to \citet{carriero2021corrigendum}, our approach includes the contemporaneous values of the endogenous variables and is thus not order invariant. In the  appendix we show that the results are robust to different orderings of the countries in $\bm y_t$, implying only negligible empirical differences.
  
Equation (\ref{eq: irga_eq}) is a simple regression model which regresses $y_{ij,t}$ on the lags of $\bm y_{it}$, the lags of the other countries' endogenous variables in $\bm z_{it}$, the contemporaneous values of the preceding $j-1$ variables domestic variables in country $i$ as well as the contemporaneous values of the preceding $i-1$ countries. Our approach builds on the notion that $\bm x_{it}$ is more important in explaining $y_{ij,t}$ than all other quantities in (\ref{eq: irga_eq}). 
  
To simplify notation, let  $\tilde{\bm z}_{ij,t} = (\bm z'_{it}, y_{i1,t}, \dots, y_{ij-1,t}, \bm y'_{1t}, \dots, \bm y'_{i-1 t})'$ denote a $K_{ij} (= K_{other} + j-1 + (i-1)M)$ vector which stores the international quantities (both lagged and contemporaneously) as well as the time $t$ values of the endogenous variables up to equation $i$. Moreover, we stack the corresponding regression coefficients in a $K_{ij}$ vector $ \tilde{\bm B}_{ij, \bullet} = (\bm B'_{ij, \bullet}, \bm u'_{ij, \bullet})'$. Notice that  $K_{ij}$ is much larger than $k$ which implies that $\tilde{\bm B}_{ij, \bullet}$ is difficult to estimate for large $M$, $N$ and $p$. We can rewrite (\ref{eq: irga_eq}) in full-data form by stacking the $T$ observations into vectors to obtain the PVAR equations which define the likelihood function of our model:
\begin{equation}
        \bm y_{ij} = \bm x_{i} \bm A_{ij, \bullet} + \tilde{\bm z}_{ij} \tilde{\bm B}_{ij, \bullet} + \bm \varepsilon_{ij}  \Leftrightarrow  \bm y_{ij} \sim \mathcal{N}({\bm x}_i \bm A_{ij, \bullet} + \tilde{\bm z}_{ij} \tilde{\bm B}_{ij, \bullet}, \sigma^2_{\varepsilon, ij} \bm I_T),
        \label{PVAR}
\end{equation}
with $\bm x_i$ being a $T \times k$ matrix where $k=Mp$ with $t^{th}$ row given by $\bm x'_{it}$. $\tilde{\bm z}_{ij}$ is a $T \times K_{ij}$ matrix with $t^{th}$ row  $\tilde{\bm z}'_{ijt}$. This equation is a regression model which discriminates between a high-dimensional set of predictors related to covariances and international quantities in $\tilde{\bm z}_{ij}$ and a low-dimensional set of domestic quantities in $\bm x_i$.

\subsection{The prior}\label{sec:priors}
The methods developed in this paper apply for any prior which has a hierarchical Gaussian form and thus leads to a full conditional posterior which is Gaussian. This is due to the fact that IRGA methods exploit the property that Gaussian distributions are invariant to rotations. A popular class of priors which has this form is the class of Gaussian global-local shrinkage priors. These can be represented as scale mixtures of Gaussians.\footnote{\cite{triplegamma} provides a taxonomy of  a range of priors in this class and discusses their properties.} 

At a general level, consider the $j^{th}$ coefficient in a model, $\phi_j$. A global-local shrinkage prior can be written as:
\begin{equation*}
\phi_j \sim \mathcal{N}(0, \psi^2_j \lambda^2), \quad \psi_j \sim f, \quad \lambda \sim g,
\end{equation*}
where $f$ and $g$ are mixing densities and many different choices for them have been proposed. In a global-local shrinkage prior, $\lambda$ controls  global shrinkage (common to all parameters). Having global shrinkage has often been found useful in Bayesian VARs (e.g., the Minnesota prior has a global shrinkage parameter) to reduce over-fitting concerns.\footnote{In Appendix \ref{sec: posterior}, we show how to implement a hierarchical Minnesota-type prior within this framework.} $\psi_j$ does local shrinkage (specific to the $j^{th}$ parameter). That is, if $\psi_j$ is estimated to be close to zero then $\phi_j$ is shrunk to be close to zero. 

Suitably chosen mixing densities $f$ and $g$ result in a wide range of popular shrinkage priors such as the LASSO \citep{lasso}, Normal-Gamma \citep{griffin2010inference}, Dirichlet-Laplace \citep{bhattacharya2015dirichlet} or the Horseshoe \citep{horseshoe}. Due to its empirical success and ease of implementation, we use the Horseshoe prior. It  
takes the form:
\begin{equation*}
\lambda \sim \mathcal{C}^+(0, 1),  \quad \psi_j \sim \mathcal{C}^+(0, 1),
\end{equation*}
whereby $\mathcal{C}^+$ denotes the half-Cauchy distribution. \cite{MS2016} show that the Horseshoe can be equivalently stated in terms of inverse Gamma distributions using suitable auxiliary variables. Specifically, 
\begin{align*}
\psi^2_{j}|\nu_{j}\sim \mathcal{G}^{-1}\left(1/2,1/\nu_{j}\right), \quad \lambda^2| \xi \sim \mathcal{G}^{-1}\left(1/2,1/\xi\right),\quad 
\nu_{j},\xi \sim \mathcal{G}^{-1}\left(1/2,1\right),
\end{align*}
with $\mathcal{G}^{-1}$ denoting the inverse Gamma distribution and $\psi_j$, $\xi$ being auxiliary parameters.   These auxiliary parameters are merely used to simplify posterior inference.  

Up to this point, we have assumed that all parameters of the model are forced to zero through a single global shrinkage parameter $\lambda$. However, in the PVAR model we have different sets of parameters, countries, and variable types (equations within a given country). Specifying a single shrinkage parameter would imply that global shrinkage is symmetric across these different dimensions, a rather restrictive assumption.  In the PVAR, we will assume that the $\lambda$ does not shrink \textit{all} parameters towards zero but is specified to differ across countries, equations and types of parameters. This implies that for each coefficient vector $\bm A_{ij, \bullet}$ and $\tilde{\bm B}_{ij, \bullet}$ we will replace $\lambda$ with $\lambda_{A,ij}$ and $\lambda_{B,ij}$, respectively. In other words, each equation will have its own global shrinkage parameters and there will be two of them: one for own country coefficients and one for other country and contemporaneous coefficients. 

Since $\tilde{\bm B}_{ij, \bullet}$ includes the covariance parameters as well, our discussion suggests that we use a  Horseshoe also on the off-diagonal elements of $\bm U$. This implies that our prior  allows for detecting whether static interdependencies across countries and variables are present and, if not, introduces shrinkage. We choose independent weakly informative inverse Gamma prior on the variances collected on the main diagonal of $\bm{H}$. In particular, $\sigma_{\epsilon,ij}^2\sim\mathcal{G}^{-1}(a_\sigma,b_\sigma)$ with $a_\sigma=b_\sigma=0.01$ for all countries and equations.

\section{Approximate Bayesian inference in PVARs}\label{sec:approxIRGA}
In this section we provide a framework that is capable of estimating huge PVAR models at reasonable computational cost. The next sub-section introduces IRGA to the PVAR case and provides some information on how posterior simulation can be carried out. Since this approach relies on approximating certain regions of the parameter space we then discuss the theoretical properties of the approximation.

\subsection{Posterior computation using integrated rotated Gaussian approximations}
As written in (\ref{PVAR}), the PVAR simply involves $MN$ regression models. Bayesian  MCMC methods for posterior and predictive inference in the regression model using a Horseshoe prior are well-established \citep[see][]{MS2016}. In theory, we could simply use such methods with our PVAR. However, the problem is that MCMC methods are simply too slow for dealing with the high-dimensional parameter spaces that arise with PVARs. The main source of this high-dimensionality is that $\tilde{\bm B}_{ij, \bullet}$ (the coefficients on contemporaneous and other country variables in the equation for the $j^{th}$ variable for country $i$) potentially contains tens of thousands of coefficients. The matrix of own country coefficients, $\bm A_{ij, \bullet}$, is much smaller. 
 
One empirical regularity often found in multi-country data sets is that own country effects are usually more important than other country effects. Hence, the literature sometimes  sets $\bm B_i$ equal to a zero matrix to rule out dynamic interdependencies. This, however, could translate into a mis-specified model. In this paper, we allow the data to speak about the degree of sparsity in $\tilde{\bm B}_{ij, \bullet}$. A second empirical regularity is that the time series in $\tilde{\bm z}_{ij}$ often display substantial co-movement \citep{kose2003international}. A potential solution would be to extract a low number of principal components from $\tilde{\bm z}_{ij}$. Another solution, which we adopt here, builds on the notion that if elements in $\tilde{\bm z}_{ij}$ are very similar to each other, it might pay off to not include all of them and effectively control for collinearity. This is also consistent with $\tilde{\bm B}_{ij, \bullet}$ being very sparse.

By contrast, $\bm A_{ij, \bullet}$ is likely non-sparse. This consideration motivates the way we implement IRGA with the PVAR. The general idea of IRGA is to use MCMC methods on important (low-dimensional) parameters but compute a fast approximation of the posterior for other (high-dimensional) parameters of less importance. In the PVAR, we consider $\bm A_{ij, \bullet}$ as the important parameters and $\tilde{\bm B}_{ij, \bullet}$ the less important ones.\footnote{It is worth noting that other choices of important and less important coefficients are possible. For instance, if it is felt likely that the U.S. is a dominant unit, then coefficients on U.S. variables could always be treated as important even for countries other than the U.S. The key consideration is that the number of parameters in $\bm A_{ij, \bullet}$ should not be too large so as to prevent practical use of MCMC methods on it.}  
 
We now provide details of how we implement IRGA in the PVAR. Let $\bm Q_i$ be the $T \times T$ rotation matrix obtained from the QR-decomposition of $\bm x_i$ and partition it as $\bm Q_i = (\bm Q_{i1}, \bm Q_{i2})$ with $\bm Q_{i1}$ being  $T \times k$ and $\bm Q_{i2}$ being $T \times (T-k)$.
 
Multiplying the equation for the $j^{th}$ variable in country $i$ by $\bm Q_i$ and exploiting the rotation invariance of the Gaussian distribution yields an equivalent representation of (\ref{PVAR}):
    \begin{align}
        \bm Q'_{i1} \bm y_{ij} &\sim \mathcal{N}(\bm Q'_{i1}   \bm x_{i} \bm A_{ij, \bullet} + \bm Q'_{i1} \bm \tilde{\bm z}_{ij} \bm \tilde{\bm B}_{ij, \bullet}, \sigma^2_{\varepsilon, ij} \bm I_{k}), \label{eq: Q_1} \\
        \bm Q'_{i2} \bm y_{ij} &\sim \mathcal{N}( \bm Q_{i2}' \tilde{\bm z}_{ij} \bm \tilde{\bm B}_{ij, \bullet}, \sigma^2_{\varepsilon, ij} \bm I_{T-k}),\label{eq: Q_2}
    \end{align}
The second equation follows since $\bm Q_{i2}' {\bm x}_{i} = \bm 0$. Note that $\bm A_{ij, \bullet}$ does not appear in it. This gives rise to a computational strategy which estimates $\tilde{\bm B}_{ij, \bullet}$ independently of $\bm A_{ij, \bullet}$. IRGA involves calculating the posteriors based on the two likelihood functions defined by (\ref{eq: Q_1}) and (\ref{eq: Q_2}). An approximate posterior for the (high-dimensional) $\tilde{\bm B}_{ij, \bullet}$ and $\sigma^2_{\varepsilon, ij}$ is obtained using  (\ref{eq: Q_2}). Conditional on this approximate posterior $\hat{p}(\tilde{\bm B}_{ij, \bullet}|\bm Q'_{i2} \bm y_{ij})$, the posterior of $\bm A_{ij, \bullet}$ is obtained using MCMC based on (\ref{eq: Q_1}).

Any Gaussian approximation can be used for $\hat{p}(\tilde{\bm B}_{ij, \bullet}|\bm Q'_{i2} \bm y_{ij})$. We use vector approximate message passing (VAMP). Our choice of VAMP is driven by its scalability in huge dimensions and the fact that recent papers in machine learning and econometrics \citep[see, e.g.,][]{korobilis2019high} have shown that it works extremely well for forecasting purposes. The specific implementation of the VAMP algorithm is the one proposed in  \cite{rangan2019vector} and  details are given in Appendix \ref{sec:VAMP}.\footnote{We have also experimented with variational Bayes to approximate $\hat{p}(\tilde{\bm B}_{ij, \bullet}|\bm Q'_{i2} \bm y_{ij})$. However, computation of the expected lower bound on evidence in these dimensions becomes computationally and numerically cumbersome. In addition, in our experiments the VAMP-based algorithm led to more precise predictions of the model.} The result is a Gaussian approximation: $\mathcal{N}(\overline{\bm B}_{ij, \bullet}, \overline{\bm V}_{ij, \bullet})$. 
 
Rewriting (\ref{eq: Q_1}) and plugging in the approximate moments of $\hat{p}(\tilde{\bm B}_{ij, \bullet}|\bm Q'_{i2} \bm y)$ yields:
    \begin{equation*}
        \bm Q'_{i1}(\bm y_{ij} - \tilde{\bm z}_{ij} \overline{\bm B}_{ij, \bullet}) \sim \mathcal{N}(\bm Q_{i1}' \bm x_{i} \bm A_{ij, \bullet}, \bm Q'_{i1} \tilde{\bm z}_{ij} \overline{\bm V}_{ij, \bullet} \tilde{\bm z}'_{ij} \bm Q_{i1} + \sigma^2_{\varepsilon, ij} \bm I_{k}).
    \end{equation*}
 This gives us a Gaussian likelihood for a regression model which can be combined with any (conditionally) Gaussian prior on $\bm A_{ij, \bullet}$ leading to a textbook form for the posterior of $\bm A_{ij, \bullet}$ which can be estimated using MCMC methods. Additional details on the full conditional posterior distributions and how we approximate the error variances as well as the hyperparameters of the prior are given in Appendix \ref{sec: posterior}.
 
 \subsection{Approximation accuracy}
In this sub-section we briefly discuss why a Gaussian approximation is reasonable and how this approximation impacts the posterior distribution of $\bm A_{ij, \bullet}$. Intuitively, the accuracy of the approximate posterior of the domestic coefficients $\hat{p}(\bm A_{ij, \bullet}|\bm y_{ij})$ depends on the goodness of the approximation to ${p}(\tilde{\bm B}_{ij, \bullet}|\bm Q'_{i2} \bm y_{ij})$. 

To investigate this relationship more formally, let $\text{KL}(p(a) || \hat{p}(a))$ denote the Kullback-Leibler (KL) divergence between an exact  and an approximating distribution. \cite{irga} show that the expected (with respect to the conditional distribution of $\bm y_{ij}$ given $\bm Q'_{i2}\bm y_{ij})$ KL divergence between $p(\bm A_{ij, \bullet}|\bm y_{ij})$ and $\hat{p}(\bm A_{ij, \bullet}|\bm y_{ij})$ is bounded from above by the approximation error to ${p}(\tilde{\bm B}_{ij, \bullet}|\bm Q'_{i2} \bm y_{ij})$. 

Formally, \citet[][Theorem 1]{irga} establish a link between the approximation quality of $\hat{p}(\tilde{\bm B}_{ij, \bullet}|\bm Q'_{i2} \bm y_{ij})$ and how this impacts the approximate full conditional posterior $\hat{p}(\bm A_{ij, \bullet}|\bm y_{ij})$:
\begin{align}
     \mathbb{E}[\text{KL}(p(\bm A_{ij, \bullet}|\bm y_{ij}) ~&||~ \hat{p}(\bm A_{ij, \bullet}|\bm y_{ij})) | \bm Q'_{i2} \bm y_{ij} ] \le \nonumber \\
     &\text{KL}({p}(\bm Q'_{i1} \tilde{\bm z}_{ij} \tilde{\bm B}_{ij, \bullet}|\bm Q'_{i2} \bm y_{ij}) \times \mathcal{N}_{\sigma_{ij}}~||~\hat{p}(\bm Q'_{i1} \tilde{\bm z}_{ij}\tilde{\bm B}_{ij, \bullet}|\bm Q'_{i2}\bm y_{ij}) \times \mathcal{N}_{\sigma_{ij}}), \label{eq: theorem1}
\end{align}
with $\mathcal{N}_{\sigma_{ij}} = \mathcal{N}(0, \sigma^2_{\varepsilon, ij} \bm I_k)$. This equation has three main implications. First, if the Gaussian approximation $\hat{p}(\tilde{\bm B}_{ij, \bullet}|\bm Q'_{i2} \bm y_{ij})$ is close to the exact full conditional posterior  ${p}(\tilde{\bm B}_{ij, \bullet}|\bm Q'_{i2} \bm y_{ij})$ the corresponding conditional distribution $\hat{p}(\bm A_{ij, \bullet}|\bm y_{ij})$ will be close to $p(\bm A_{ij, \bullet}|\bm y_{ij})$.  Second, the prior on $\bm A_{ij, \bullet}$ does not impact the error bound. Third, it does not depend on concentration properties around the true value of $\tilde{\bm B}_{ij,\bullet}$ which makes the result relevant for settings with $T$ being small.

In the next step, we justify our Gaussian approximation. Under certain mild assumptions on ${p}(\tilde{\bm B}_{ij, \bullet}|\bm Q'_{i2} \bm y_{ij})$ and $\tilde{\bm z}_{ij}$ and if $k \ll K_{ij}$, a multivariate central limit theorem implies that $\bm Q'_{i1} \tilde{\bm z}_{ij} \tilde{\bm B}_{ij, \bullet}$ is close to a Gaussian distribution \citep[see][]{diaconis1984asymptotics} even if elements in $\tilde{\bm B}_{ij, \bullet}$ are non-Gaussian.  This  motivates a Gaussian approximation to ${p}(\tilde{\bm B}_{ij, \bullet}|\bm Q'_{i2} \bm y_{ij})$. \cite{irga}, in Theorem 2, show that the expected KL divergence (with respect to $\bm Q'_{i2} \tilde{\bm z}_{ij}$) between the actual full conditional and the  Gaussian approximation is bounded  by two constants $\varpi_1$ and $\varpi_2$:\footnote{The constants $\varpi_1$ and $\varpi_2$ take a complicated form. To avoid introducing additional notation we summarize their main properties here. Precise definitions can be found in \cite{irga}.}

\small\begin{equation}
 \mathbb{E} \left[ \text{KL}({p}(\bm Q'_{i1} \tilde{\bm z}_{ij} \tilde{\bm B}_{ij, \bullet}|\bm Q'_{i2} \bm y_{ij}, \bm Q'_{i2}\tilde{\bm z}_{ij}) \times \mathcal{N}_{\sigma_{ ij}}~||~\hat{p}(\bm Q'_{i1} \tilde{\bm z}_{ij}\tilde{\bm B}_{ij, \bullet}|\bm Q'_{i2}\bm y_{ij}, \bm Q'_{i2}\tilde{\bm z}_{ij}) \times \mathcal{N}_{\sigma_{ij}})\right]\le \varpi_1 + \varpi_2, \label{eq:theorem2}
\end{equation}\normalsize
$\varpi_1$ depends on the concentration properties of $p(\tilde{\bm B}_{ij, \bullet}|\bm Q'_{i2} \bm y_{ij})$ around its mean and  $\varpi_2$ on the average correlation between the elements in $\tilde{\bm B}_{ij, \bullet}$. 

If the posterior covariance of $\tilde{\bm B}_{ij, \bullet}$ is small relative to $\sigma^2_{\varepsilon, ij}$ and when $k \ll K_{ij}$, $\varpi_1$ approaches zero. The second constant measures approximation errors between the first two moments of ${p}(\tilde{\bm B}_{ij, \bullet}|\bm Q'_{i2} \bm y_{ij})$  and the approximating density. This quantity  depends on the posterior covariance of the approximating density to the posterior of $\tilde{\bm B}_{ij, \bullet}$ and thus the error bound can be small even if the approximate variance-covariance matrix differs sharply from the true covariance of the posterior of $\tilde{\bm B}_{ij, \bullet}$. This finding also has important implications for our estimates of $\bm A_{ij, \bullet}$ since  (\ref{eq: theorem1}) and (\ref{eq:theorem2}) can be combined to arrive at an upper bound for the approximation error to the posterior distribution of $\bm A_{ij, \bullet}$.

 \section{Illustration using synthetic data}\label{sec: synth}
The theoretical discussion in the previous section builds on certain assumptions about the rows of $\tilde{\bm z}_{ij}$ and the correlation properties of the posterior of  ${p}(\tilde{\bm B}_{ij, \bullet}|\bm Q'_{i2} \bm y_{ij})$.  In this  section, we will use synthetic data to illustrate how our approach performs under a realistic data generating process (DGP). 

Our DGP assumes that $N=10, K=2, n=20$ and $T=500$ and features a single lag. It is given by:
\begin{align*}
    \bm y_{t} &=\hat{ \bm \Gamma}_1 \bm y_{t-1} + \bm \epsilon_{t}, \quad  \bm \epsilon_{t} \sim \mathcal{N}(\bm 0_{20}, \bm U \bm H \bm U'),
\end{align*}
where $ \bm H = \sigma^2_\varepsilon \times  \bm I_{20}$ and $u_{ij} \sim \mathcal{N}(0, \sigma^2_\varepsilon/10) \text{ for } i=2, \dots, n; j=1, \dots, i-1$. The matrix $\hat{ \bm \Gamma}_1$ is obtained as follows. The blocks referring to the domestic coefficients are centred around the same mean vector and we add Gaussian shocks:
\begin{align*}
  \text{vec}{(\bm \Gamma_{i1})} = \text{vec}\left[\begin{pmatrix} 0.8 & 0.2 \\ 0.3 & 0.6
  \end{pmatrix}\right] + \sigma_\beta \bm \eta_t, \quad \bm \eta_t \sim \mathcal{N}(\bm 0_4, \bm I_4),
\end{align*}
for all countries $i$. We simulate the elements in $\bm \Xi_{i}$ from $\mathcal{N}(0, \sigma_\beta^2)$. To obtain sparsity in $\bm \Xi_i$ we randomly zero out elements such that we have around $60$ percent zeroes in $\bm \Xi_i$. The initial value of $\bm y_t$ is sampled from a multivariate zero-mean Gaussian with variance $0.01$. To analyze how the goodness of our approximation changes with different measurement errors and cross-country heterogeneity, we consider $\sigma_\varepsilon, \sigma_\beta \in \{0.01, 0.025, 0.05\}$. 

The main question is whether our IRGA-based approach yields estimates of the domestic coefficients which are close to the ones obtained from exact methods.   Since exact methods become prohibitively slow in truly large  data sets we make this comparison operational by simulating from a moderately large DGP and estimate a PVAR model with a Horseshoe prior and without using IRGA (i.e., all coefficients are estimated using MCMC). For both models we include $p=2$ lags of $\bm y_t$. 

To avoid mixing up  approximation errors arising from using IRGA to any errors in estimates that come from our equation-by-equation estimation approach, both the exact and approximate models are estimated based on the same (correct) ordering and in what follows we compare the coefficients $\bm A_{i1}$ and $\bm B_i$ to their true (implied) values.
\begin{table}[h!]
\caption{Simulation evidence: IRGA-PVAR versus PVAR estimated through MCMC.}\label{tab: sim}
\centering
  \rowcolors{2}{gray!25}{white}
\begin{tabular}{lrrr}
    \rowcolor{gray!50}
  \toprule
  \backslashbox{$\sigma_\varepsilon$}{$\sigma_\beta$} & $0.01$ & $0.025$ & $0.05$  \\ 
  \midrule
   $0.01$  & 1.99 & 1.33 & 1.06 \\ 
   & 1.96 & 1.29 & 1.00  \\ 
   $0.025$ & 1.19 & 1.08 & 1.04  \\ 
   & 1.19 & 1.07 & 0.97  \\ 
   $0.05$  & 1.11 & 1.09 & 1.07  \\ 
   & 1.14 & 1.10 & 1.03  \\ 
     \bottomrule
\end{tabular}\\
 \begin{minipage}{6cm}~\\
\tiny \textit{Notes}:
The table shows the relative Mean Absolute Error (MAE) ratios between our IRGA-PVAR and a PVAR estimated using MCMC techniques for different values of $\sigma^{\text{*}}_\varepsilon$ and $\sigma^{\text{*}}_\beta$. The numbers are averages across $100$ draws from the DGP. The gray shaded rows refer to the MSE ratios for the full coefficient vector (i.e., including both international and domestic coefficients) while the white rows include MSE ratios for the domestic coefficients only.
\end{minipage}%
\end{table}

Table \ref{tab: sim} shows averages of relative mean absolute error (MAE) ratios between the IRGA-PVAR and the PVAR estimated through MCMC across $100$ replications of the DGP. The white rows refer to MAE ratios for the domestic coefficients whereas the gray shaded rows denote MAE ratios for all regression coefficients. For both models we use the posterior median as our point estimator.

When we consider the results for the domestic coefficients (i.e., the white rows) we find a great deal of relative MAEs close to $1$ (except for cases in which both $\sigma_\beta$ and $\sigma_\varepsilon$ are very small). In principle, when we fix a given row and consider increasing values of $\sigma_\beta$  we find that the relative MAEs approach unity.  This is consistent with the theoretical predictions in the previous section and shows that, at least when the posterior mean is considered, approximation accuracy of our IRGA-based approach increases with the ratio $\sigma_\beta / \sigma_\varepsilon$.  When we fix a given column, as long as $\sigma_\varepsilon$ is not too small, we find no discernible differences across values of $\sigma_\beta$.

Turning to the relative MAEs  across all coefficients  reveals that our approximate method also yields estimates of $\bm B_i$ which are competitive to the ones obtained using MCMC.  In fact, for $\sigma_\beta=0.05$ we find that the mean estimates essentially equal to the ones obtained from the MCMC-based method.

\begin{figure}
    \centering
    \includegraphics[scale=0.41]{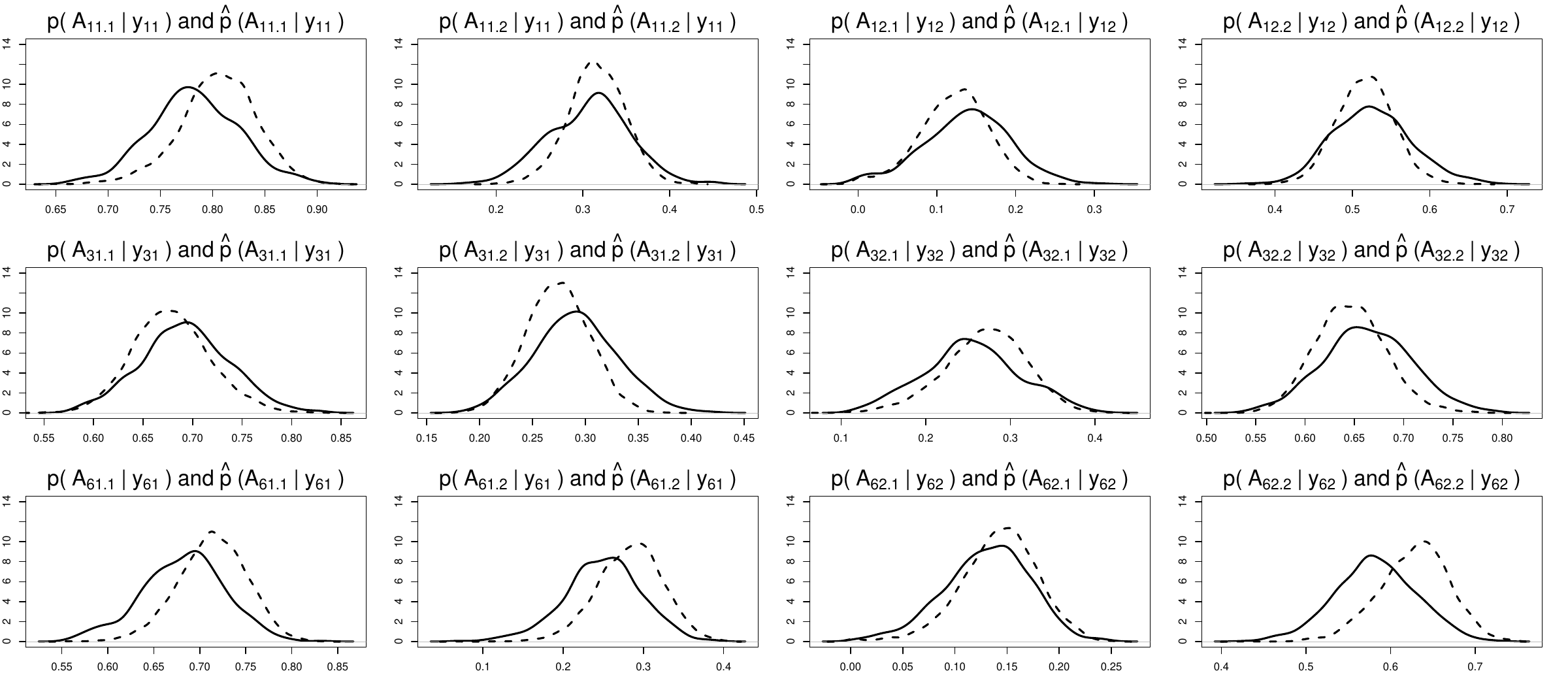}
    \caption{Marginal posterior distribution of the domestic coefficients in $\bm A_{ij, \bullet}$ for three countries.}
    \label{fig:sim_approx}\vspace{-0.25cm}
     \caption*{\footnotesize{\textit{Notes:} 
     $A_{ij, s}$ denotes the $s^{th}$ element of $\bm A_{ij, \bullet}$ for $s=1, \dots, k$.
     The solid black line refers to the approximate posterior $\hat{p}(A_{ij, s}| \bm y_{ij})$ while the dashed black line denotes the posterior distribution ${p}(A_{ij, s}| \bm y_{ij})$ obtained through estimating the model using MCMC.}}
\end{figure}

To illustrate our approach using a single draw from the DGP for $\sigma_\beta = \sigma_\varepsilon=0.05$, \autoref{fig:sim_approx} shows the marginal posterior distributions of the domestic coefficients for our IRGA-based PVAR (in solid black) and the MCMC-based estimates (in dashed black). This figure shows that in most cases, posterior distributions are similar. Especially when we focus on the mean/median we observe only small differences across coefficients (with some few cases suggesting a larger disagreement between MCMC and approximate estimates). When we focus on the higher moments of the marginal distributions we find similar variances, tail behavior and skewness properties. This small discussion has shown that, at least when synthetic data is considered, our approach yields reasonable estimates. 

\section{A huge model of the world economy}\label{sec: emp_work}
In this section we develop a huge-dimensional model of the world economy. The model is used to forecast output (measured by industrial production), inflation, long-term interest rates and stock prices for a large panel of countries. We moreover analyze the properties of the forecasts using novel stochastic block models as well as Diebold-Yilmaz (DY) spillover indices.

\subsection{Data}\label{sec:data}
We have collected macroeconomic and financial data from the OECD's short-term indicator data base. The data are monthly and span the period from 2001m2 to 2019m12. In principle, we have a panel of 18 series for 38 OECD countries but not all variables are available with the same country coverage (see Table \ref{tbl:data}). In total we have $487$ variables in our PVAR. 

The series fall into three categories: macroeconomic, financial and leading indicators. For macroeconomic variables, we consider measures of economic activity (industrial production growth, the output gap and the unemployment rate), export and import growth, consumer and producer price inflation as well as changes in earnings in the manufacturing sector. 

Financial data cover short- and long-term interest rates (overnight, 3-months money market rates and long-term government bond yields), changes in stock prices and broad money growth. For euro area countries, we include the 3-months euribor as a measure of short-term interest rates. 

In addition to this rather standard macro-financial data set, we gathered data on confidence/sentiment and short-term leading indicators. These comprise the OECD's leading indicator, which is a constructed measure to provide early signals of turning points in business cycles,\footnote{This measure shows fluctuations of an economy's activity around its long-term trend (i.e., potential output). The main difference to the output gap measure we employ is that the leading indicator is amplitude adjusted. For more information, see \href{http://www.oecd.org/sdd/leading-indicators/oecdcompositeleadingindicatorsreferenceturningpointsandcomponentseries.htm}{oecd.org/sdd/leading-indicators/oecdcompositeleadingindicatorsreferenceturningpointsandcomponentseries.htm}.}  measures of manufacturers' and consumers' confidence, as well as changes in passenger car registration and newly permitted dwellings. Confidence measures potentially contain additional information to predict economic activity \citep{Batchelor1998, Ludvigson2004} as do car registrations and new dwellings for consumer expenditures. Changes or growth rates refer to either year-on-year or month-on-month growth rates. For a detailed overview, see Table \ref{tbl:data}.

\begin{table}[htbp]
    \centering    \caption{Data description.}\label{tbl:data}\small
    \begin{tabular}{llcr}
    \toprule
         \textbf{OECD code}	&	\textbf{Description}	&	\textbf{Trans.} & \textbf{Coverage (percent)}	\\
         \midrule
         \multicolumn{4}{l}{\textit{Macroeconomic data}} \\
PRINTO01	&	Industrial production, s.a.	&	1	&	81.6	\\
LORSGPRT	&	GDP, ratio to trend (output gap)	&	0	&	92.1	\\
LRHUTTTT	&	Harmonized unemployment rate: all persons, s.a.	&	0	&	76.3	\\
XTEXVA01	&	Exports in goods, s.a.	&	1	&	100.0	\\
XTIMVA01	&	Imports in goods, s.a.	&	1	&	100.0	\\
CPALTT01	&	Consumer prices: all items	&	1	&	92.1	\\
PIEAMP01	&	Producer prices - Manufacturing	&	1	&	60.5	\\
LCEAMN01	&	Hourly earnings: manufacturing, s.a. &	2	&	39.5	\\
\midrule
\multicolumn{4}{l}{\textit{Financial data}}				\\
IRSTCI01	&	Overnight interbank rate	&	0	&	96.3	\\
IR3TIB01	&	3 month interbank rate	&	0	&	92.6	\\
IRLTLT01	&	Long-term interest rate	&	0	&	78.9	\\
SPASTT01	&	Share prices	&	2	&	100.0	\\
MABMM301	&	Broad money, s.a.	&	2	&	28.9	\\
\midrule
\multicolumn{4}{l}{\textit{Leading indicators}}					\\
LOLITOAA	&	Leading indicator, amplitude adjusted	&	0	&	92.1	\\
BSCICP02	&	Manufacturing confidence indicator, s.a.	&	0	&	78.9	\\
CSCICP02	&	Consumer confidence indicator, s.a.	&	0	&	65.8	\\
SLRTCR03	&	Passenger car registrations, s.a.	&	2	&	28.9	\\
ODCNPI03	&	Permits issued for dwellings, s.a.	&	2	&	31.6	\\
\bottomrule 
    \end{tabular}
    \begin{minipage}{16cm}~\\
\scriptsize \textit{Notes}: All data are from the OECD's short-term indicator database, accessed via dbnomics (\url{https://db.nomics.world/}). Transformations refer to no-transformation (0), year-on-year growth rates (1) and month-on-month growth rates (2). Country coverage of each variable in percent. The sample consists of  Austria, Australia, Belgium, Brazil, Canada, Switzerland, China, Czechia, Germany, Denmark, Estonia, Spain, Finland, France, Great Britain, Greece, Hungary, Indonesia, Ireland, Israel, Italy, Japan, Korea, Lithuania, Latvia, Mexico, Netherlands, Norway, New Zealand, Poland, Portugal, Russia, Sweden, Slovenia, Slovakia, Turkey, U.S.A, and South Africa.
\end{minipage}%
\end{table}

\subsection{Design of the forecasting exercise and competing models}
We carry out a forecasting exercise comparing an unrestricted PVAR with a Horseshoe prior to a range of alternatives.  In what follows, we focus on predictions for four \textit{target variables}: consumer price inflation (labeled Infl. in the tables and figures), industrial production (Ind. prod.),  stock returns (Equities) and long-term interest rates (LT-IR).

Our forecasting design is recursive, implying that we use the data from 2001m2 to 2006m12 as an initial estimation period. We use data through 2006m12 to produce one-month up to twelve-months-ahead forecast distributions. This initial estimation period is then extended by data for 2007m1 and forecast densities for 2007m2 (up to 2008m1) are constructed. We repeat this procedure until we reach the end of the hold-out period.

To compare point forecasts across models we use MAEs. Since this disregards higher-order moments of the predictive distribution we also use Log Predictive Likelihoods (LPLs) to compare the density forecast performance of these alternatives.

The set of competing models is chosen not only to reflect a variety of popular approaches, but also to be computationally practical. In particular, alongside the proposed PVAR-IRGA approach, we consider the following models. First, to assess the role of allowing for cross-country spillovers, we estimate single-country Bayesian VARs (BVARs) that rule out both static and dynamic interdependencies. These are estimated country-by-country. Second, to compare our approach to other models incorporating international information, we consider two types of specifications. The first is a factor-augmented VAR (FAVAR-10) model, which augments the single-country BVARs with $10$ factors extracted from the non-domestic country variables. This procedure serves to obtain a lower dimensional representation of the international information set. As a second option to include international information, we use a Bayesian GVAR.

To ensure that differences in forecast performance are not driven by the respective priors on the VAR coefficients, we estimate all  models with Horseshoe priors and set the number of lags equal to two. As a robustness check we also repeat the forecasting exercise replacing the Horseshoe prior with a conventional Minnesota prior for each of the PVAR-IRGA, BVAR, FAVAR-10 and the GVAR. Further specification details and results are given in Appendices \ref{sec: posterior}, \ref{app:othermodels} and \ref{sec:resMN}.

Other than PVAR-IRGA, all of the models are estimated using exact MCMC methods. We stress that approaches which involve using MCMC methods for an unrestricted $487$ dimensional VAR would simply be computationally impractical. Computation times (average estimation time per model over all periods in the hold-out) for doing the pseudo real time forecasting exercise are provided in Table \ref{tab:times}. It can be seen that PVAR-IRGA is substantially faster than any of the competing approaches, even though all of the latter are much more parsimonious models. Notice that estimating the PVAR-IRGA is even faster than estimating a set of $N$ country-specific VARs of the same size. This is because MCMC sampling of the domestic coefficients based on the IRGA posterior is faster since it only relies on a part of the likelihood function to form the conditional posterior distributions. 
\begin{table*}[ht]
\caption{Estimation time in minutes.}\label{tab:times}\vspace*{-1.5em}
\small
\begin{center}
\begin{threeparttable}
\begin{tabular*}{\textwidth}{@{\extracolsep{\fill}}rrrr}
  \toprule
 \textbf{BVAR} & \textbf{FAVAR-10} & \textbf{GVAR} & \textbf{PVAR-IRGA} \\ 
  \midrule
 235.4 min & 663.7 min & 566.4 min & \textbf{113.4 min} \\ 
   \bottomrule
\end{tabular*}
\begin{tablenotes}[para,flushleft]
\scriptsize{\textit{Notes}: BVAR and FAVAR-10 are estimated one country at a time. The indicated time marks the time required for estimating all forecasts for the full system. GVAR and PVAR-IRGA estimation times are based on joint estimation of the multi-country system.}
\end{tablenotes}
\end{threeparttable}
\end{center}
\end{table*}

 \subsection{Summary of forecasting results}
The results of our forecasting exercise are summarized in Table \ref{tb2}. It presents absolute values of MAEs and LPLs for our PVAR-IRGA approach (rows shaded in red) for two forecast horizons. Results for the other approaches are benchmarked relative to these. To be precise, for MAEs we take ratios relative to PVAR-IRGA (with numbers exceeding unity implying that the PVAR improves upon the competitors), for LPLs we take differences relative to PVAR-IRGA (with values below zero suggesting that the PVAR is outperforming the respective model). {The numbers in the table are GDP-weighted averages of country-specific LPLs computed by using GDP in 2015 U.S. dollars averaged over the period 2002 to 2019. To provide a rough gauge of model performance across countries, the numbers in parentheses represent the percentage of countries in which a given model performs best in absolute terms.}

The most important finding is that PVAR-IRGA works -- it produces sensible forecasts quickly. Bayesian estimation of huge dimensional PVARs has been made possible through the use of IRGA methods.  {The other main finding is that (with some exceptions) PVAR-IRGA works well and is highly competitive with competing approaches. These improvements are limited for point forecasts but sometimes very pronounced for LPLs which measure density forecast performance. For short-term forecasts of  equity returns in particular, PVAR-IRGA is producing strong improvements in LPLs while its performance is slightly weaker for the remaining variables under consideration.}

{When we focus on higher-order forecasts the relative performance of PVAR-IRGA improves. While we observe that predictive accuracy is deteriorating for equities at the twelve-months-ahead horizon, forecasts of industrial production and inflation improve considerably. For the latter two variables, the PVAR-IRGA is the single best performing model.}

BVAR is the only alternative that does not allow for any cross-country spillovers. On average, it forecasts fairly well for both forecast horizons. Improvements, however, are more more pronounced at the one-month-ahead horizon. For twelve-months-ahead, we find that taking cross-country linkages into account helps forecast accuracy. This suggests that for short-run forecasts, cross-country spillovers are not that strong (or at least do not significantly help in predicting output, inflation and long-term rates). 

The fact that the single country BVARs improve upon the GVAR indicates that having a smaller-sized sparse model seems to be more important than taking into account cross-country linkages for improving forecasts. The  good performance of our PVAR-IRGA suggests that the Horseshoe prior successfully strikes a balance between exploiting cross-country information and sparsity. Since the model is unrestricted it also allows the data to speak about the precise form of cross-country linkages.

The statements in the preceding paragraphs are based on an examination of LPLs. Analyzing MAEs reveals similar patterns, but to a weaker extent. This indicates that the benefits of unrestricted modeling of the high-dimensional PVAR offers some benefits in terms of point forecast performance, but the benefits are larger for density forecasts. 

\begin{table}[!tbp]
\scriptsize
\caption{Summary of Forecast Exercise, all models with Horseshoe prior.}\label{tb2}\vspace*{-1.5em}
\begin{center}
\begin{tabular}{llcrrrrcrrrr}
\toprule
\multicolumn{1}{l}{\bfseries }&\multicolumn{1}{c}{\bfseries }&\multicolumn{1}{c}{\bfseries }&\multicolumn{4}{c}{\bfseries MAE}&\multicolumn{1}{c}{\bfseries }&\multicolumn{4}{c}{\bfseries LPS}\tabularnewline
\cline{4-7} \cline{9-12}
\multicolumn{1}{l}{}&\multicolumn{1}{c}{Model}&\multicolumn{1}{c}{}&\multicolumn{1}{c}{Equities}&\multicolumn{1}{c}{Ind. prod.}&\multicolumn{1}{c}{LT-IR}&\multicolumn{1}{c}{Infl.}&\multicolumn{1}{c}{}&\multicolumn{1}{c}{Equities}&\multicolumn{1}{c}{Ind. prod.}&\multicolumn{1}{c}{LT-IR}&\multicolumn{1}{c}{Infl.}\tabularnewline
\midrule
{\itshape h=1}&&&&&&&&&&&\tabularnewline
\shadeRow   ~~&   BVAR&   &   1.008&   0.988&   0.927&   1.000&   &     -4.686&      2.538&     13.855&     -0.503\tabularnewline
\shadeRow   ~~&   &   &   (31.6)&   ( 9.7)&   (36.7)&   (22.9)&   &   (13.2)&   (12.9)&   (13.3)&   (14.3)\tabularnewline
   ~~&   FAVAR-10&   &   1.004&   0.981&   0.923&   0.994&   &     -4.511&      4.022&     14.664&      0.713\tabularnewline
   ~~&   &   &   (23.7)&   (51.6)&   (53.3)&   (60.0)&   &   ( 2.6)&   (51.6)&   (66.7)&   (45.7)\tabularnewline
\shadeBench   ~~&   PVAR-IRGA&   &   0.825&   0.410&   0.274&   0.386&   &   -177.799&    -39.815&     36.181&    -13.387\tabularnewline
\shadeBench   ~~&   &   &   (44.7)&   (25.8)&   (10.0)&   (17.1)&   &   (68.4)&   (25.8)&   (20.0)&   (40.0)\tabularnewline
   ~~&   GVAR&   &   1.051&   1.092&   1.090&   1.199&   &    -12.541&    -21.086&    -42.623&    -44.480\tabularnewline
   ~~&   &   &   ( 0.0)&   (12.9)&   ( 0.0)&   ( 0.0)&   &   (15.8)&   ( 9.7)&   ( 0.0)&   ( 0.0)\tabularnewline
\midrule
{\itshape h=12}&&&&&&&&&&&\tabularnewline
\shadeRow   ~~&   BVAR&   &   0.984&   1.081&   0.912&   1.020&   &      9.251&    -10.936&     16.181&     -4.867\tabularnewline
\shadeRow   ~~&   &   &   (39.5)&   (32.3)&   (40.0)&   (40.0)&   &   (42.1)&   (25.8)&   (50.0)&   (34.3)\tabularnewline
   ~~&   FAVAR-10&   &   0.987&   1.089&   0.896&   1.020&   &      5.545&    -15.878&     14.098&     -0.372\tabularnewline
   ~~&   &   &   (28.9)&   ( 3.2)&   (16.7)&   (11.4)&   &   ( 5.3)&   ( 6.5)&   (20.0)&   (22.9)\tabularnewline
\shadeBench   ~~&   PVAR-IRGA&   &   0.875&   0.664&   0.570&   0.759&   &   -204.541&   -149.824&   -104.962&   -174.713\tabularnewline
\shadeBench   ~~&   &   &   (28.9)&   (48.4)&   (36.7)&   (45.7)&   &   (47.4)&   (61.3)&   (30.0)&   (40.0)\tabularnewline
   ~~&   GVAR&   &   1.048&   1.072&   0.981&   1.135&   &     -1.333&    -13.215&    -15.904&    -30.407\tabularnewline
   ~~&   &   &   ( 2.6)&   (16.1)&   ( 6.7)&   ( 2.9)&   &   ( 5.3)&   ( 6.5)&   ( 0.0)&   ( 2.9)\tabularnewline
\bottomrule
\end{tabular}
\end{center}
\vspace*{-1em}\tiny{\textit{Notes:} GDP-weighted average over countries, win percentage across countries in parentheses. Root mean squared error (RMSE) and log predictive score (LPS) relative to the benchmark. The benchmark PVAR-IRGA (shaded in red) shows actual values, all other models are in ratios to the benchmark for RMSEs and in differences for LPSs.}
\end{table}
{The discussion has focused on averages across countries (and time periods). However, it could be that a model (such as our PVAR-IRGA) yields lower average LPLs but still provides the best performance for individual countries in our sample. Considering the percentage of wins for each model corroborates the findings based on LPLs and MAEs. But it is worth emphasizing that even though the PVAR-IRGA is sometimes outperformed by simpler competitors such as BVAR, for most variables we still find a sizable fraction of wins across countries for our proposed model. In the case of one-month-ahead inflation density forecasts, this share is about 34 percent whereas it is around 47 percent for longer-run forecasts of equity returns (in which BVAR outperforms the PVAR-IRGA if we consider LPLs). These sizable shares suggest that it might be worthwhile to carefully analyze country-specific results.}

\subsection{Forecast comparison across countries and over time }
In the preceding sub-section, we compared the average (over countries and  time) forecast performance of PVAR-IRGA to various alternatives. In this sub-section, we look behind the average performance to investigate forecast performance at the country level and see how it changes over time. We do so through heatmaps of cumulative LPLs for PVAR-IRGA for the individual countries. {To assess when it pays off to allow for cross-country linkages (and because it is the strongest competitor to our approach) we benchmark our results to the LPLs of the BVAR.}

\begin{figure}
    \begin{subfigure}{0.24\linewidth}
    \caption{Infl.}
        \includegraphics[width=\linewidth]{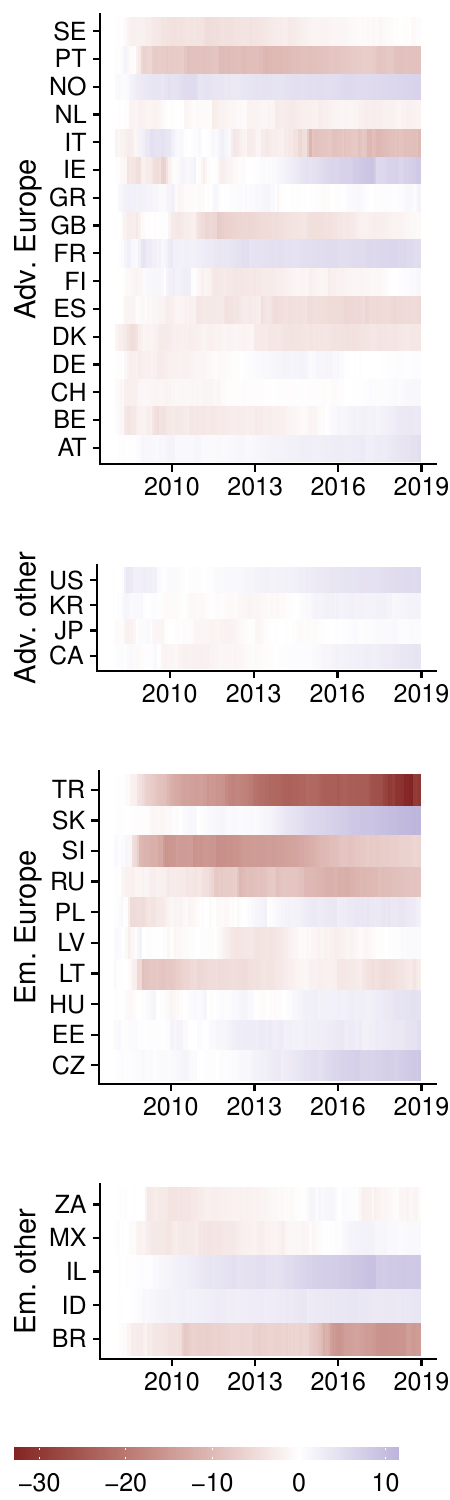}
    \end{subfigure}
    \begin{subfigure}{0.24\linewidth}
    \caption{Ind. Prod.}
        \includegraphics[width=\linewidth]{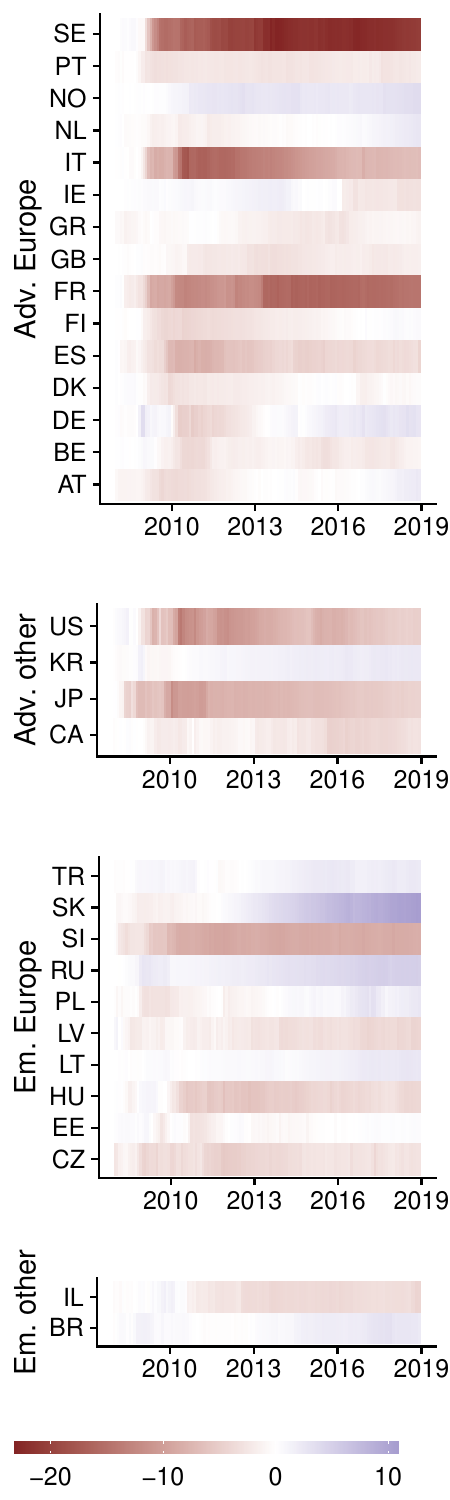}
    \end{subfigure}
    \begin{subfigure}{0.24\linewidth}
    \caption{Equities}
        \includegraphics[width=\linewidth]{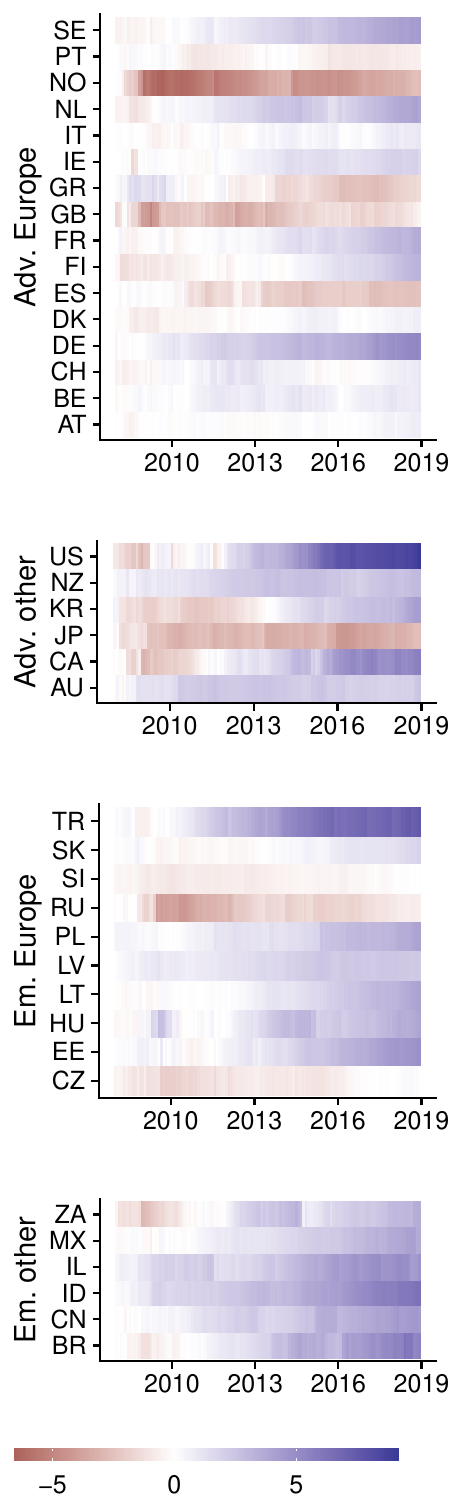}
    \end{subfigure}
    \begin{subfigure}{0.24\linewidth}
    \caption{LT-IR}
        \includegraphics[width=\linewidth]{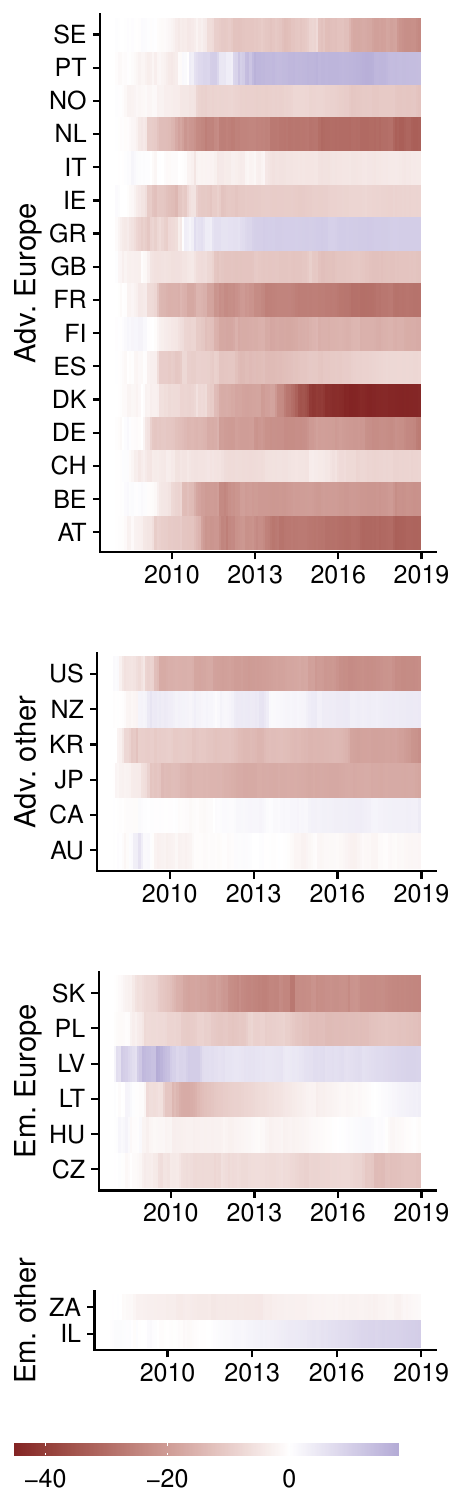}
    \end{subfigure}
    \caption{Cumulative LPS for PVAR-IRGA relative to BVAR for $h=1$ forecasts across countries and over time.}
    \label{fig:cLPS-h1}
\end{figure}

\begin{figure}
    \begin{subfigure}{0.24\linewidth}
    \caption{Infl.}
        \includegraphics[width=\linewidth]{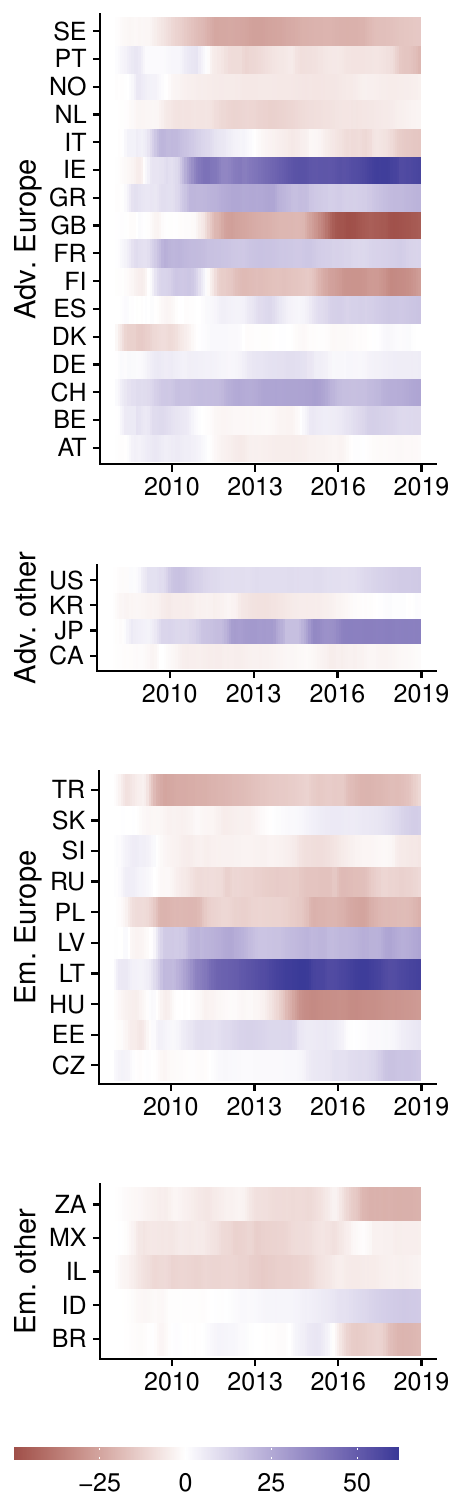}
    \end{subfigure}
    \begin{subfigure}{0.24\linewidth}
    \caption{Ind. Prod.}
        \includegraphics[width=\linewidth]{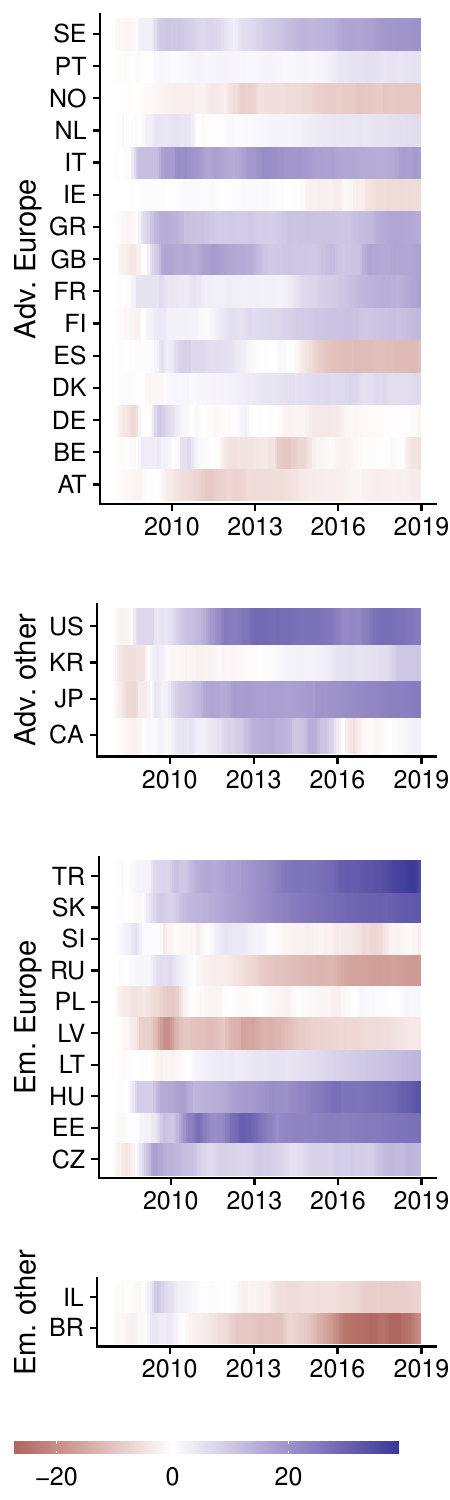}
    \end{subfigure}
    \begin{subfigure}{0.24\linewidth}
    \caption{Equities}
        \includegraphics[width=\linewidth]{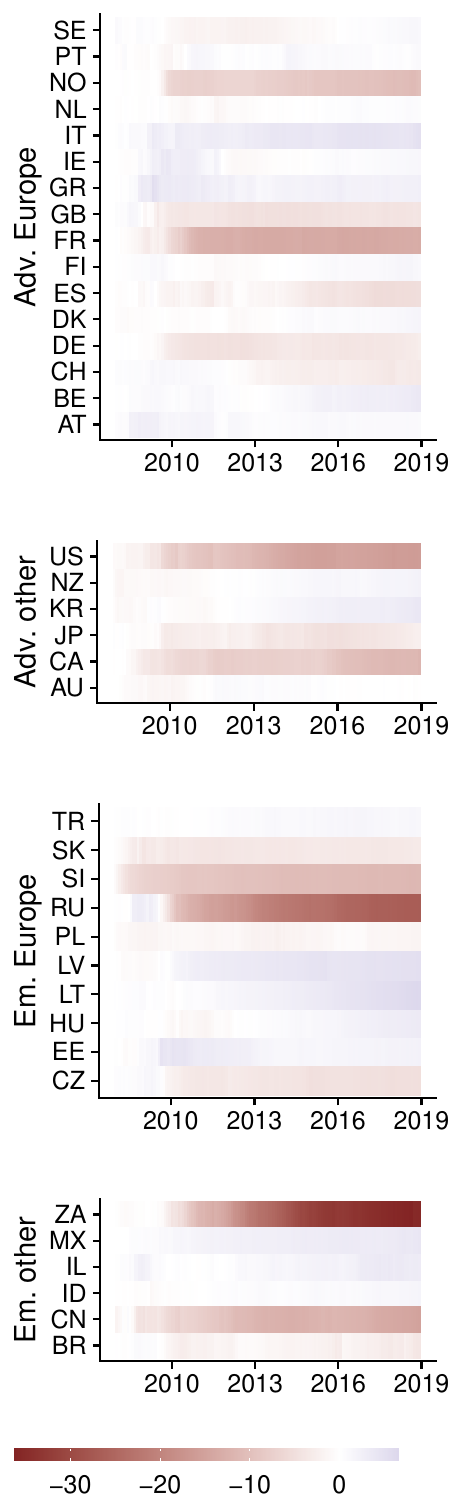}
    \end{subfigure}
    \begin{subfigure}{0.24\linewidth}
    \caption{LT-IR}
        \includegraphics[width=\linewidth]{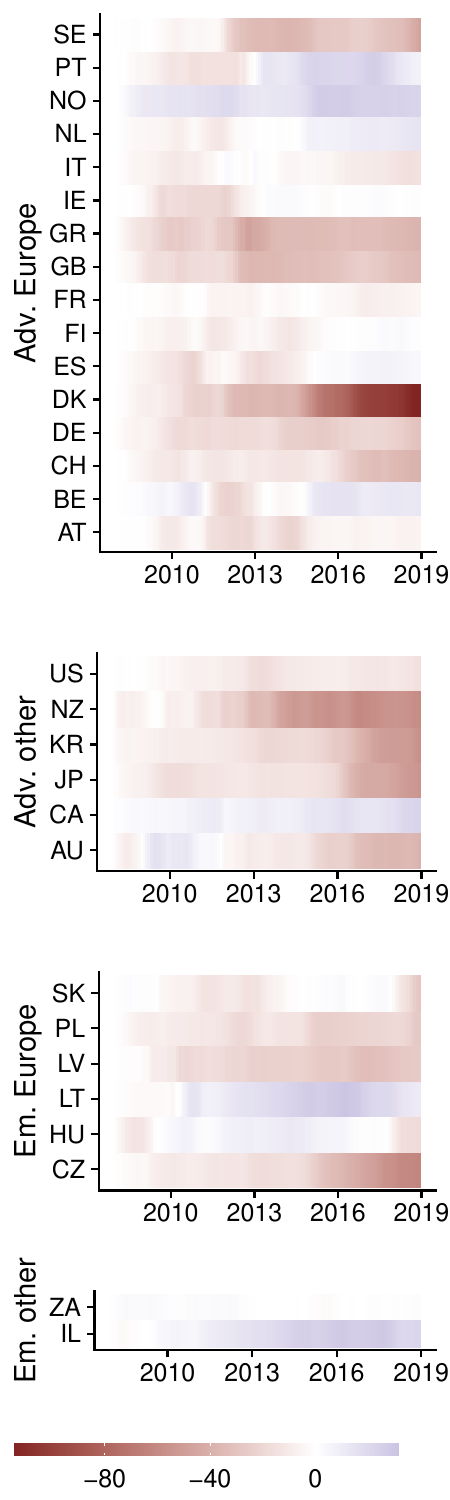}
    \end{subfigure}
    \caption{Cumulative LPS for PVAR-IRGA relative to BVAR for $h=12$ forecasts across countries and over time.}
    \label{fig:cLPS-h12}
\end{figure}

Figures \ref{fig:cLPS-h1} and \ref{fig:cLPS-h12} contain these heatmaps for the two forecast horizons. The figures are grouped into four categories: Advanced European, Emerging European, Advanced Other and Emerging Other and individual countries are labeled using ISO country codes. Intensifying shades of blue (red) indicate stronger support for PVAR-IRGA (BVAR). 

{We first focus on one-month-ahead forecasts. Starting with inflation predictions, we observe that the stronger performance of the BVAR is mainly driven by a weak performance of the PVAR-IRGA in emerging economies (most notably Brazil and Turkey, with some exceptions, such as India) while it performs best for developed economies such as the U.S., France and Ireland. This brief discussion shows why it is important to also consider results at the country level. If the researcher is interested in short-term forecasting of U.S. inflation and has to decide on one of the models we consider, focusing on overall LPLs masks the particularly strong U.S.-specific forecasting performance.}

{When we focus on industrial production we find a somewhat different pattern, with PVAR-IRGA being outperformed by the BVAR for several countries in Advanced Europe (except for Austria, Finland, Norway and the Netherlands) and some gains in several countries located in Emerging Europe (Slovakia, Russia and the Baltics). For some countries (such as Sweden and Italy) the PVAR-IRGA performs well prior to the global financial crisis. The rapid decline in output in the final half of 2008,  however, led to a deterioration in forecast performance. This is because PVAR-IRGA yields predictive distributions which are sometimes too tight and thus capturing outliers becomes increasingly difficult.}

{Turning to the results for equity returns reveals a great number of blue-colored cells. In principle, our model works well for most economies (with a slightly weaker performance for, e.g., Portugal, Norway, Slovenia and Japan). Interestingly, we also find some heterogeneity with respect to model performance over time. In the case of the U.S., for instance, our model only improves upon the BVAR from 2012 onward. We conjecture that the slightly weaker performance prior and during the financial crisis is, again, driven by too tight predictive intervals. But these tight intervals then help in predicting returns after the financial crisis, a period characterized by steady increases in U.S. stock markets.}

{The PVAR-IRGA displays the weakest performance for long-term interest rates. For some few countries (e.g., Portugal, Greece, Latvia and Israel) the PVAR performs  well.  In general, the weak performance for long-term rates is driven by the fact that these display a downward trend during the hold-out period for most countries. PVAR-IRGA captures this downward trend rather well but the predictive variance is considerably smaller than the one of the single-country BVAR. Hence, under the predictive distribution of the PVAR-IRGA, even  relatively small changes in long rates have strong effects on LPLs. The countries which depart from this general pattern (such as Portugal and Greece) feature large spikes in long-term interest rates. The PVAR captures this well and quickly adjusts the predictive variance. Since it takes slightly longer for the BVAR to adjust we conjecture that this quick increase is mostly driven by the large information set.}

{In the previous sub-section we have shown that on average, the PVAR-IRGA produces the most precise density forecasts for inflation and industrial production when the forecast horizon is increased while the performance for equity returns deteriorates. When we focus on twelve-month-ahead predictions (see \autoref{fig:cLPS-h12}) we find that the strong overall performance for inflation is mostly driven by excellent forecasts in major developed economies such as the U.S. or Japan. A similar pattern is found for industrial production. Again, we find that the PVAR-IRGA produces precise density forecasts for most developed economies located in Europe as well as the United States and Japan. However, it is also worth stressing that the PVAR also produces accurate forecasts for developing economies such as Turkey as well as several countries located in Central Eastern Europe. This strong performance is driven by more precise point forecasts but also by the fact that the predictive distributions for multi-step-ahead forecasts seem to be heavy tailed and thus make observing outliers more probable.}

{For equities and long-term interest rates we find that the PVAR performs slightly weaker than the BVAR. Especially for equities, this is driven by a particularly bad performance in the U.S. For long-term interest rates we again find that the PVAR is competitive when used to forecast long-term rates in Portugal and Israel but it appears to be outperformed in countries such as Denmark.}

{In summary, PVAR-IRGA yields precise equity return predictions for short-term forecasts and shows good performance when used to forecast inflation in major economies such as the U.S. For higher-order forecasts, the results somewhat reverse and the PVAR-IRGA works particularly well when it is used to predict industrial production and inflation. In general, and this is consistent with the findings based on average LPLs, we find that more predictive evidence in favor of cross-country spillovers increases with the forecast horizons. We will provide additional evidence on the increasing importance of cross-country spillovers in the following sub-sections.}

\subsubsection{A quantitative analysis of the point forecasts}
In the previous section we have shown that our PVAR yields highly competitive forecasts and often improves upon other single- or multi-country models. In this section, our aim is to quantitatively analyze the country-specific forecasts to investigate the role of cross-country correlations in the point predictions. For brevity, we focus on one-month-ahead forecasts for industrial production and long-term interest rates.

Since the corresponding correlation matrices are high-dimensional and thus difficult to interpret, we use techniques from network analysis \citep[see, e.g.,][]{vziberna2014blockmodeling} to search for clusters in the correlation matrices of the posterior median and the posterior standard deviation of the forecasts.\footnote{These are implemented through the R package \texttt{blockmodeling}.} Intuitively speaking, we reorder the rows and columns such that correlations between countries are grouped into $R=8$ distinct blocks.\footnote{The choice of 8 blocks is arbitrary. For our application, it yields a good balance between a too granular and a too coarse approach.} The relations (i.e., positive correlations) within a given block are maximized whereas the relations of countries within a block to economies outside of a block are much less important (or even negative).

Since this algorithm needs a correlation matrix of forecasts, we compute the initial correlation matrix based on the first 12 observations and then expand this  window until we reach the end of the hold-out period which is used to compute forecast distributions. This yields a sequence of correlation matrices which we then analyze using a stochastic block model. 

\begin{figure}
    \begin{subfigure}{.49\linewidth}
    \caption{January 2009}\vspace*{-0.5em}
        \includegraphics[width=\linewidth, trim=80 0 1 100, clip]{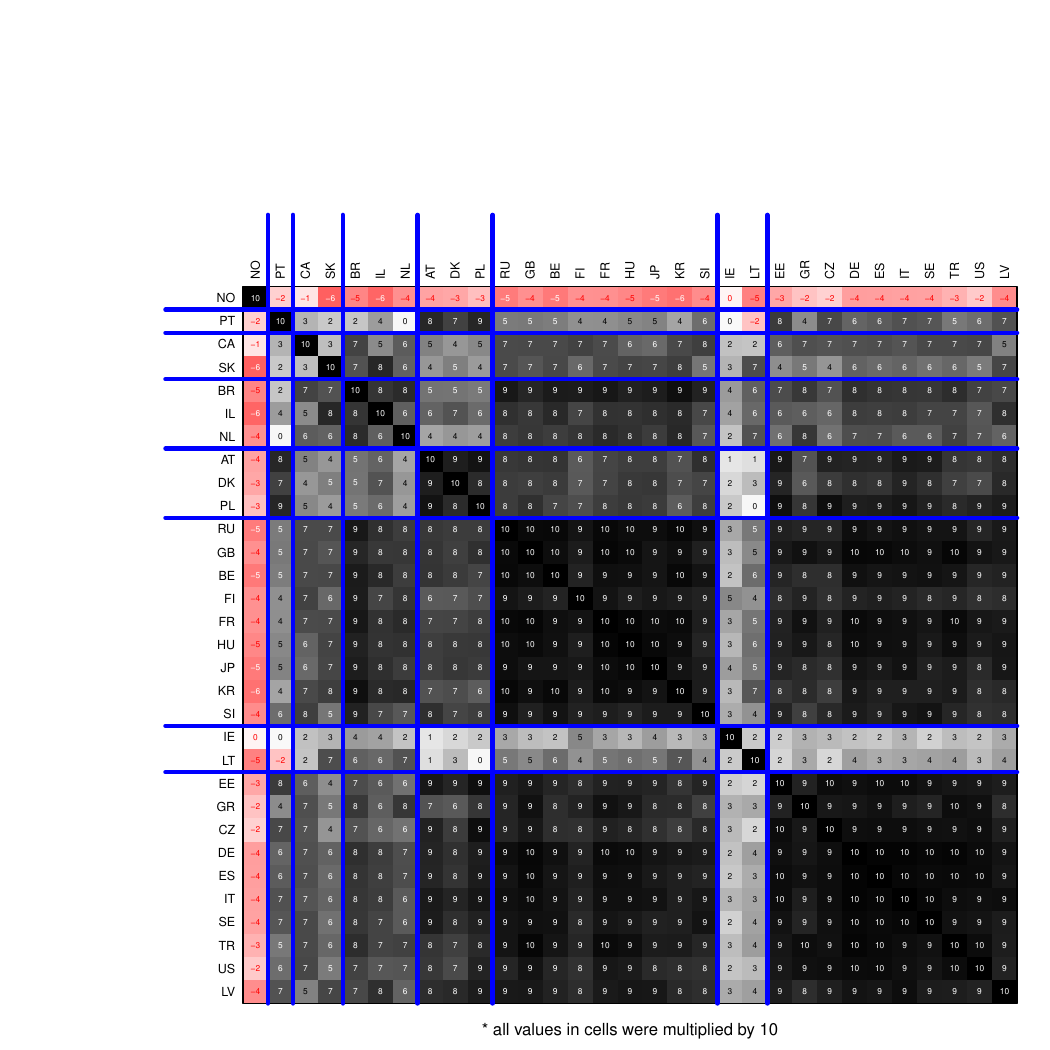}
    \end{subfigure}
    \begin{subfigure}{.49\linewidth}
    \caption{June 2012}\vspace*{-0.5em}
        \includegraphics[width=\linewidth, trim=80 0 1 100, clip]{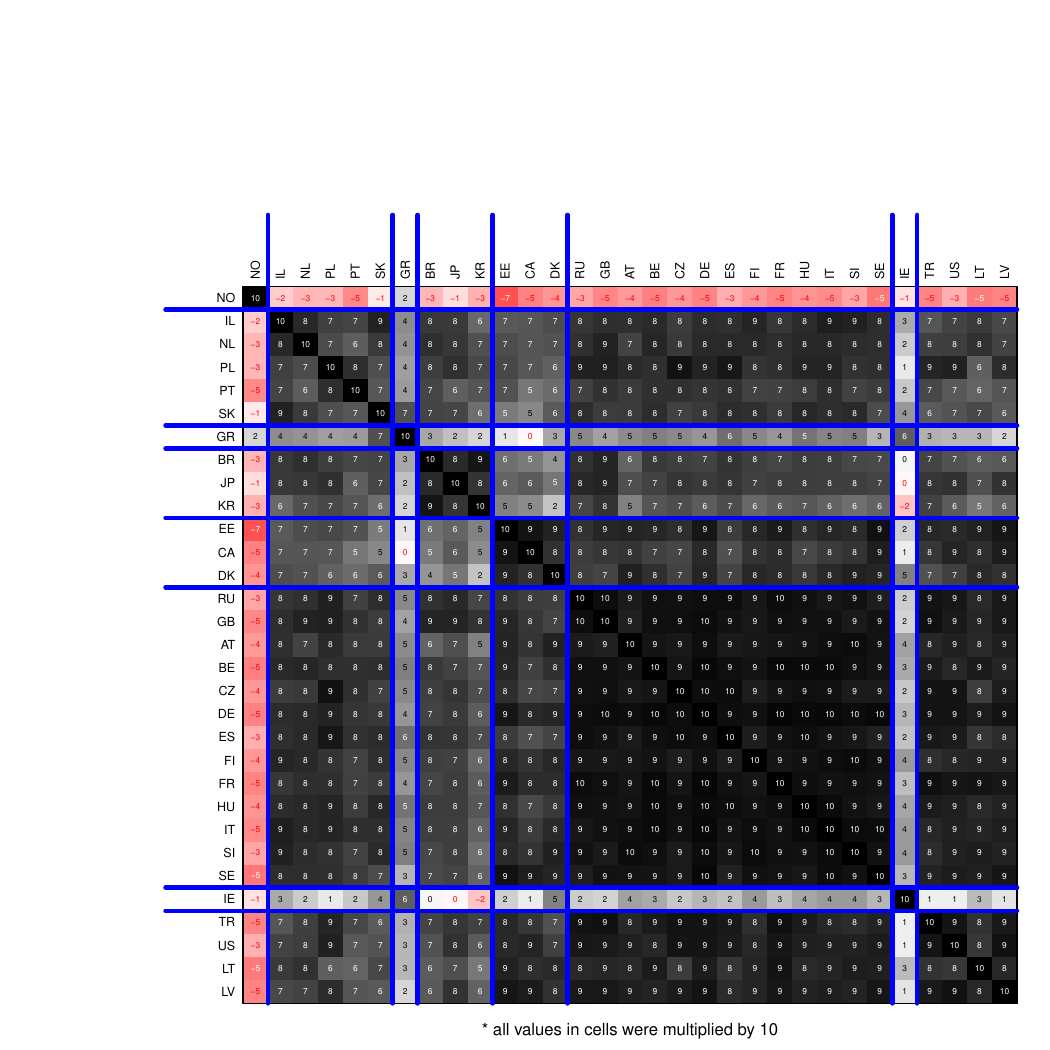}
    \end{subfigure}
    \\[1em]
    \begin{subfigure}{.49\linewidth}
    \caption{December 2019}\vspace*{-0.5em}
        \includegraphics[width=\linewidth, trim=80 0 1 100, clip]{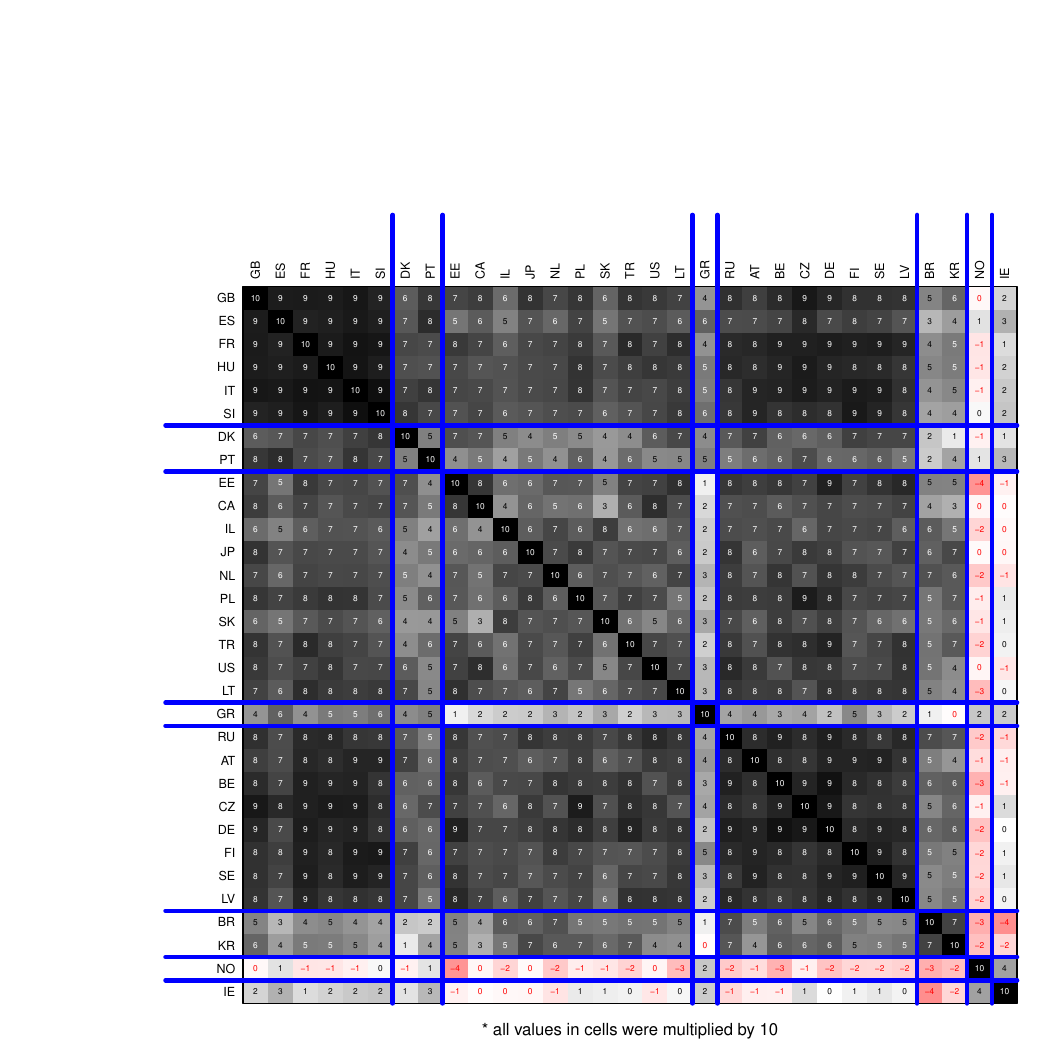}
    \end{subfigure}
    \caption{Correlation matrices of one-month-ahead forecasts for industrial production sorted using a stochastic block model.} \label{fig: stochblock_UR}
\end{figure}

In what follows, we focus on one-month-ahead forecasts for three distinct periods in our hold-out sample. First, we examine correlation structures among our forecasts for  2009m1, the onset of the global financial crisis. The second period we consider is 2012m6, the month prior to Mario Draghi's famous ``whatever it takes'' speech, which marks the height of the euro area sovereign debt crisis. The final period is 2019m12, the end of our sample. Using all available information is a natural choice to investigate how forecasts are related. Figure \ref{fig: stochblock_UR} shows the correlation matrices (multiplied by 10) for industrial production forecasts sorted using a stochastic block model. 

The figure shows the $R=8$ clusters on the main diagonal of the matrix. The first two blocks are each defined by a single country (Norway and Portugal, respectively). Forecasts between the blocks are (modestly) negatively correlated, as indicated by the red shading in the off-diagonal elements between the blocks. Considering the remaining countries, we see that forecasts for Norway are not only negatively correlated with those for Portugal, but also with the rest of the sample. As an oil-based economy, Norway was considerably less affected by the global financial crisis than the rest of the sample. Looking at the remaining economies reveals two large clusters. One of them contains forecasts for Greece, Italy, Spain and the Baltics --  countries that showed massive contractions in output during the global financial crisis. The Irish economy, which also significantly contracted during the global financial crisis, appears in another cluster. Another large cluster contains countries that were comparably less affected by the crisis. Importantly though and with the exception of Norway, all clusters are positively correlated indicated by the grey to black colour shading in Figure \ref{fig: stochblock_UR}. This reflects the global nature of the financial crisis. 

Considering  the period of the euro area sovereign debt crisis, depicted in the upper right panel of Figure \ref{fig: stochblock_UR} reveals a very similar picture: Norway stands out, and forecasts of the remaining countries are positively correlated. One big cluster emerges which covers mostly European economies and Russia -- the latter which shares strong trade ties with the European Union. Other European countries that were not affected by the sovereign debt crisis, such as the Baltics, are allocated into different blocks and Ireland again appears in a separate cluster. Turning now to one-month ahead forecasts for the end of the hold-out sample, we still find isolated countries such as Norway and Ireland, two European clusters and one large, international cluster. The latter one shows only a modest within-correlation, indicating that at the end of our sample, with the pandemic yet to fully unfold, forecasts are less strongly correlated as in periods of severe downturns. 

\begin{figure}
    \begin{subfigure}{.49\linewidth}
    \caption{January 2009}\vspace*{-0.5em}
        \includegraphics[width=\linewidth, trim=80 0 1 100, clip]{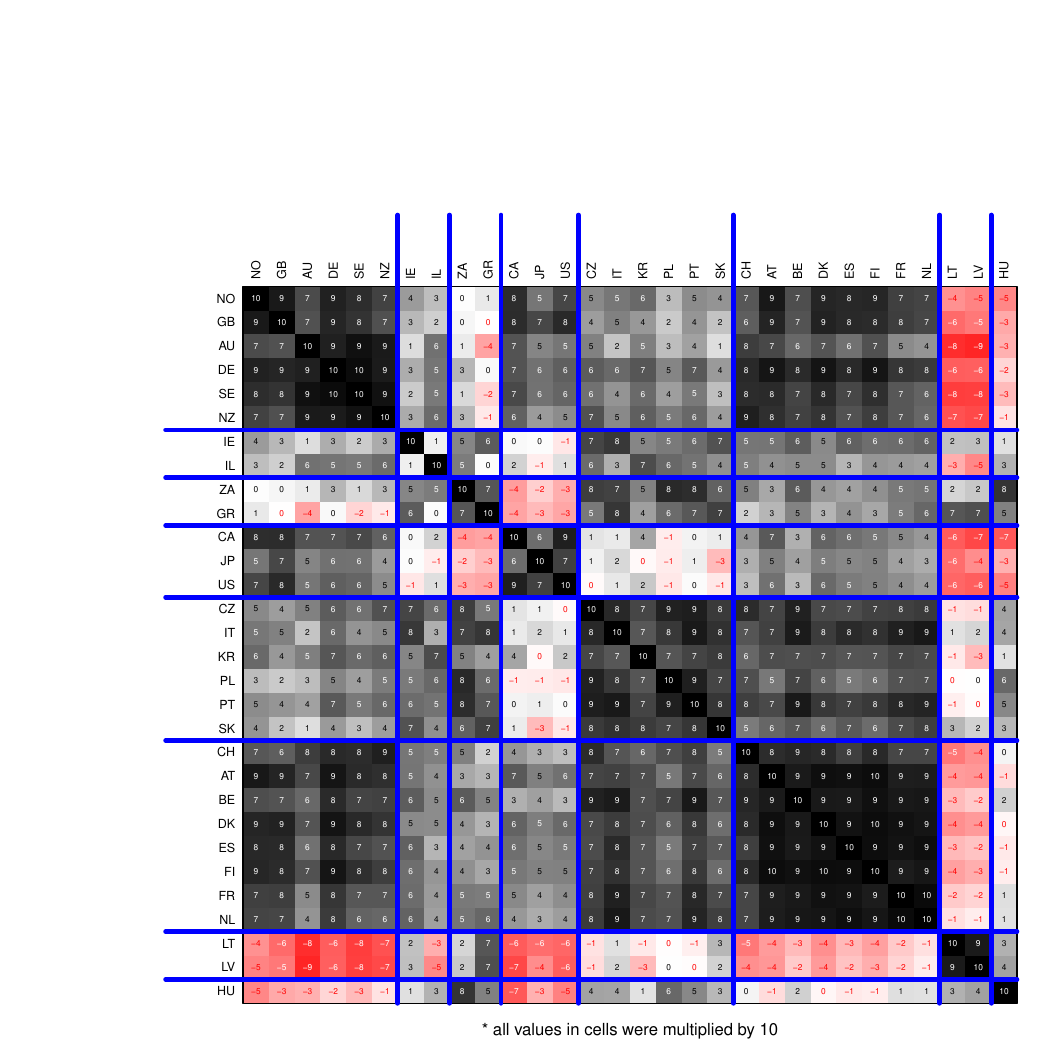}
    \end{subfigure}
    \begin{subfigure}{.49\linewidth}
    \caption{June 2012}\vspace*{-0.5em}
        \includegraphics[width=\linewidth, trim=80 0 1 100, clip]{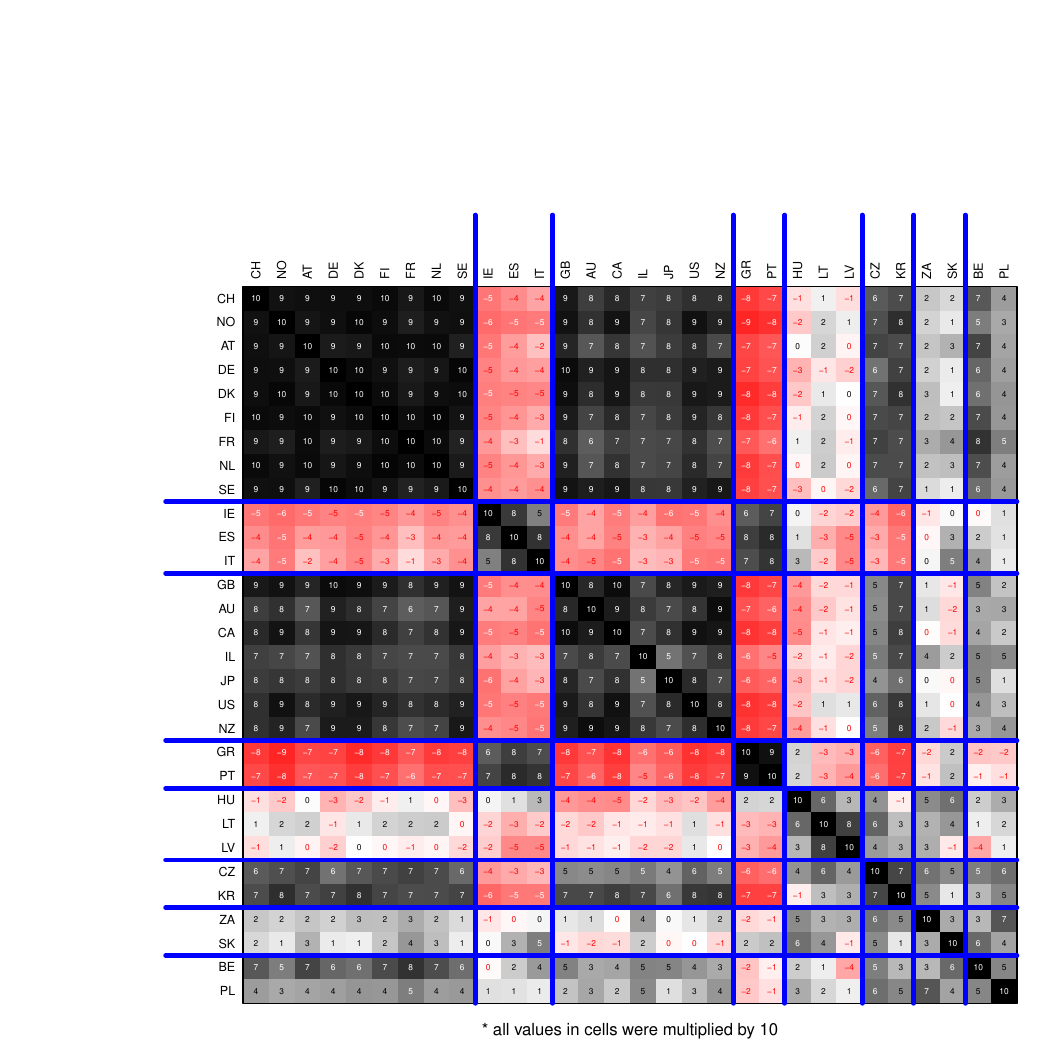}
    \end{subfigure}
    \\[1em]
    \begin{subfigure}{.49\linewidth}
    \caption{December 2019}\vspace*{-0.5em}
        \includegraphics[width=\linewidth, trim=80 0 1 100, clip]{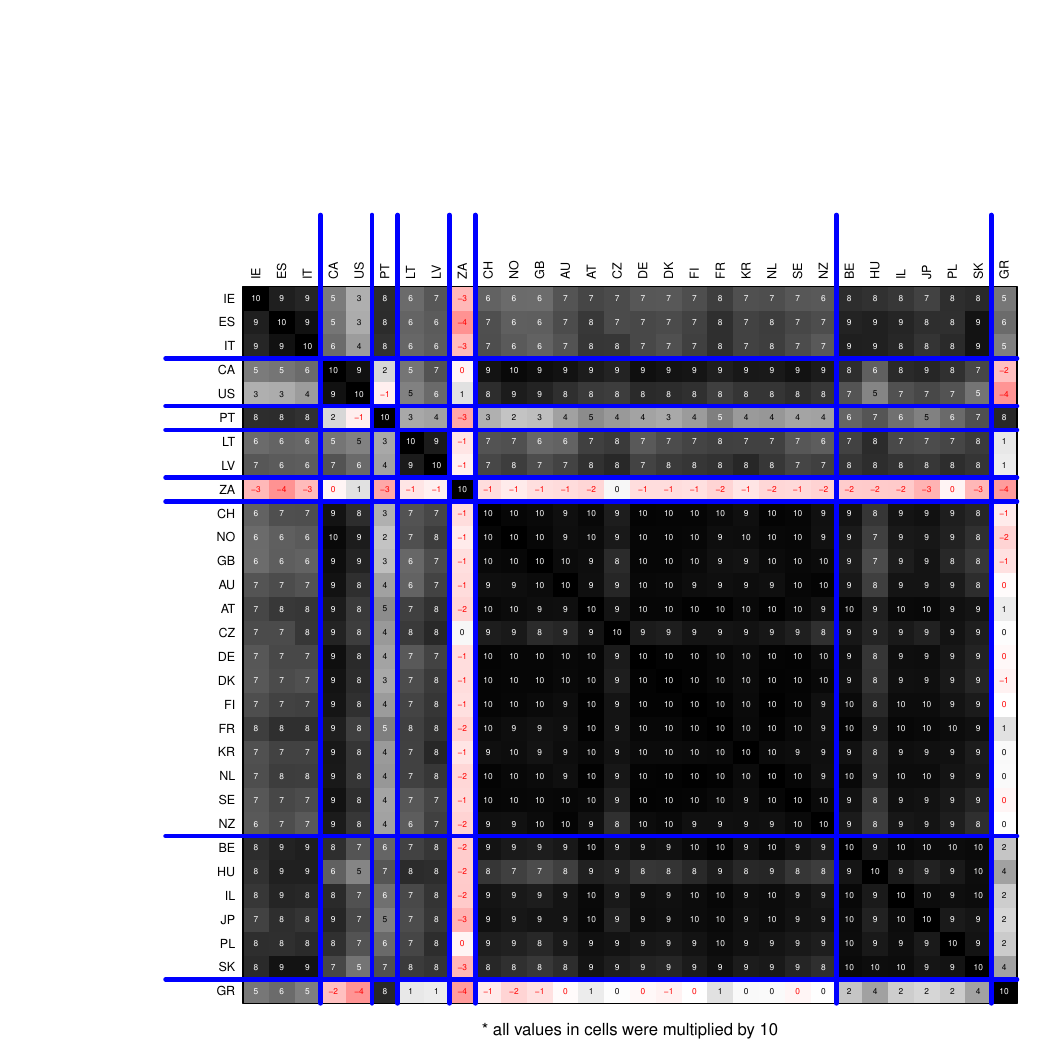}
    \end{subfigure}
    \caption{Correlation matrices of one-month-ahead forecasts for long-term interest rates sorted using a stochastic block model.} \label{fig: stochblock_ltir}
\end{figure}

Figure \ref{fig: stochblock_ltir} shows the same analysis for long-term interest rate forecasts. During the period of the global financial crisis, long-term rates spiked in countries like Greece, but have been downward-trending in countries considered as safe havens: the United States, Canada and Japan. PVAR-IRGA forecasts are consistent with these historical observations putting the aforementioned country groups into separate blocks. Two further clusters emerge: one solely consisting of European economies, while the other one contains both Advanced European and non-European economies. At the height of the sovereign debt crisis, long-term interest rates shot up for most economies but to a different extent. This is mirrored in the first four correlation blocks depicted on the main diagonal of the correlation matrix. Forecasts for crisis-stricken economies, such as Ireland, Italy and Spain are strongly correlated and appear in a separate block. Interestingly, Greece and Portugal appear in one block, but forecasts between these clusters are positively correlated. Both blocks of crisis stricken countries are clearly separated from the rest of the sample. Last and looking at the end of sample period, we see a very different picture. With the exception of South Africa and to some extent Greece, global long-term interest forecasts are very homogeneous and positively correlated

Summing up, examining correlation structures of one-month ahead forecasts revealed insights as to which extent the model is capable of mirroring correlation structures present in the data. Correlations are strongest during periods of simultaneous contractions such as witnessed during the global financial crisis. Looking at the episode of the sovereign debt crisis, which unfolded on a more regional basis, also reveals clusters but they differ from those identified during the global financial crisis. This highlights the overall flexibility of the model. For some countries forecasts are always separated from the rest of the sample (e.g., Norway as an oil-exporting economy). This implies that the PVAR-IRGA can take cross-country links into account when they are important but at the same time does not enforce them on the whole set of countries -- a flexibility which is of ample importance when dealing with large, heterogeneous cross-sections.

\subsubsection{Other evidence of cross-country spillovers}
Another way of looking at cross-country correlations is to use the Diebold-Yilmaz (DY) spillover index \citep[see][]{dy2009}. This index  distinguishes to which extent forecast error variance can be explained by its own history as opposed to effects through all other variables in the system. The latter  effects are dubbed "spillovers" and serve as a measure of overall connectivity. For instance, in this paper, we calculate the index  based on a generalized forecast error variance decomposition (GFEVD) which avoids order dependence with respect to the elements in $\bm y_t$. We report the total share of effects from non-domestic variables (spillovers from one variable to another) in the GFEVD. The calculation of the underlying GFEVD is recursive (in the same manner as was done for the forecasting exercise), and we show results for a forecast horizon of $12$ and $24$ months.  Our focus on higher-order spillovers is motivated by evidence reported in the literature on GVARs \citep[see, e.g.,][]{Feldkircher2016a} which shows that cross-country spillovers (measured through FEVDs) become sizable only after several quarters.

\begin{figure}
    \includegraphics[width=\textwidth]{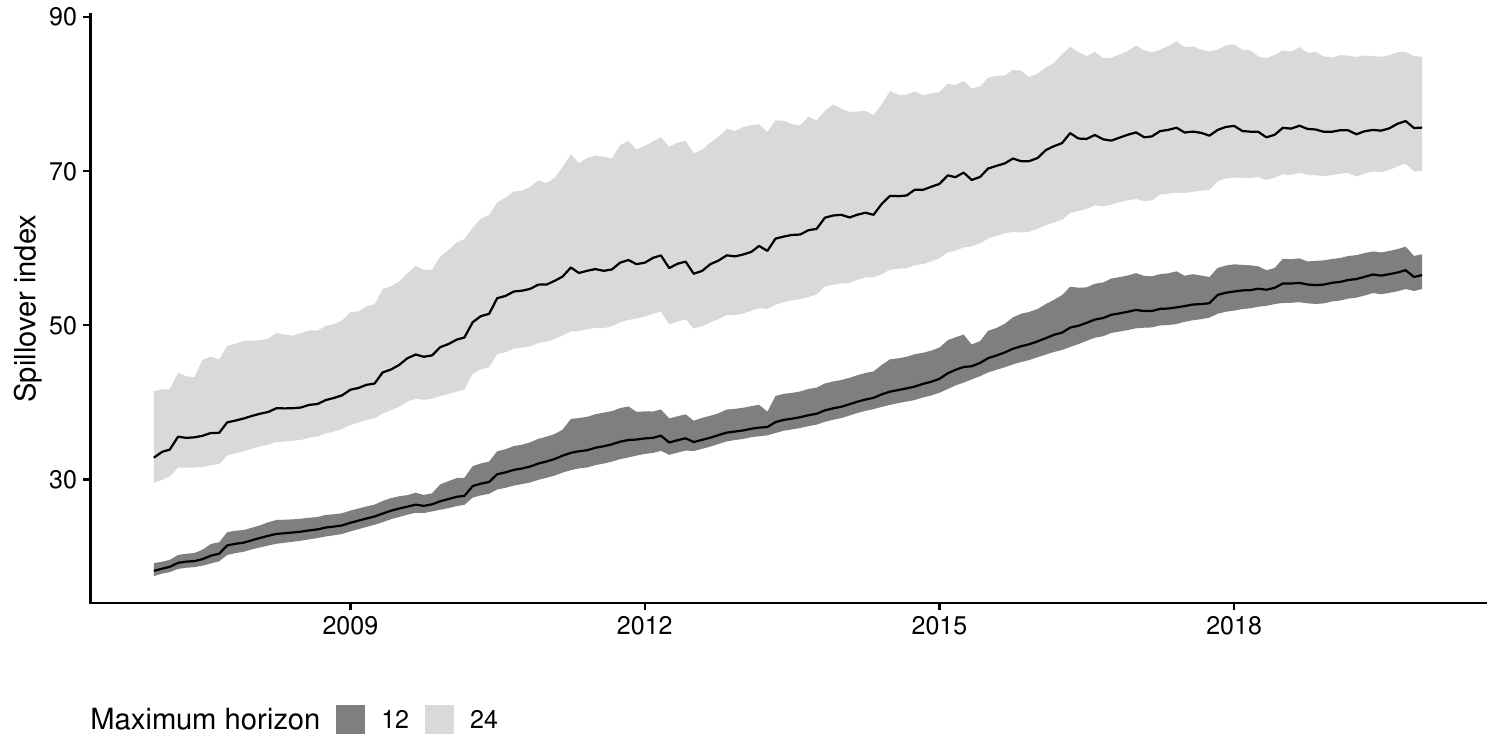}\\ 
    \caption{Diebold-Yilmaz spillover index for all countries based on decomposition of forecast error variance of the full system.}\vspace*{-1em}
    \caption*{\footnotesize{\textit{Notes:} This index indicates the share of spillovers across countries excluding spillovers to variables within a given country (summed across variable types). Estimated based on an expanding window. The solid line is the posterior median alongside the 68 percent posterior credible set.}}
    \label{DYindex}
\end{figure}
{Figure \ref{DYindex} shows the DY index over the hold-out period for both horizons. Most importantly, our results point at a sizable degree of cross-country connectedness. At the end of our sample period, the DY indices amount to about 45 at the twelve-months-ahead forecast horizon  and to 68 percent at the 24-months-ahead horizon. Investigating the DY index over time, reveals a steady increase of the index until 2016 after which the indicator levels out.\footnote{Note that consistent with the forecasting exercise, we use an expanding window to calculate the DY index which automatically introduces a certain degree of persistence. Appendix \ref{sec:resMN} provides additional empirical results for estimates using a rolling window of observations.} A particularly pronounced increase of the index can be observed during
the period of the euro area sovereign debt crisis (between 2010 and 2012). In general, economic variables tend to co-move more strongly during turbulent times \citep[see, e.g.,][]{Pham2021} -- a pattern that we also observe with our data. It is worth stressing that our estimates are surrounded by considerable posterior uncertainty, which tends to attribute more posterior mass above than below the posterior median. This behavior is even more pronounced during the period of the sovereign debt crisis.}

\begin{figure}
    \includegraphics[width=\textwidth]{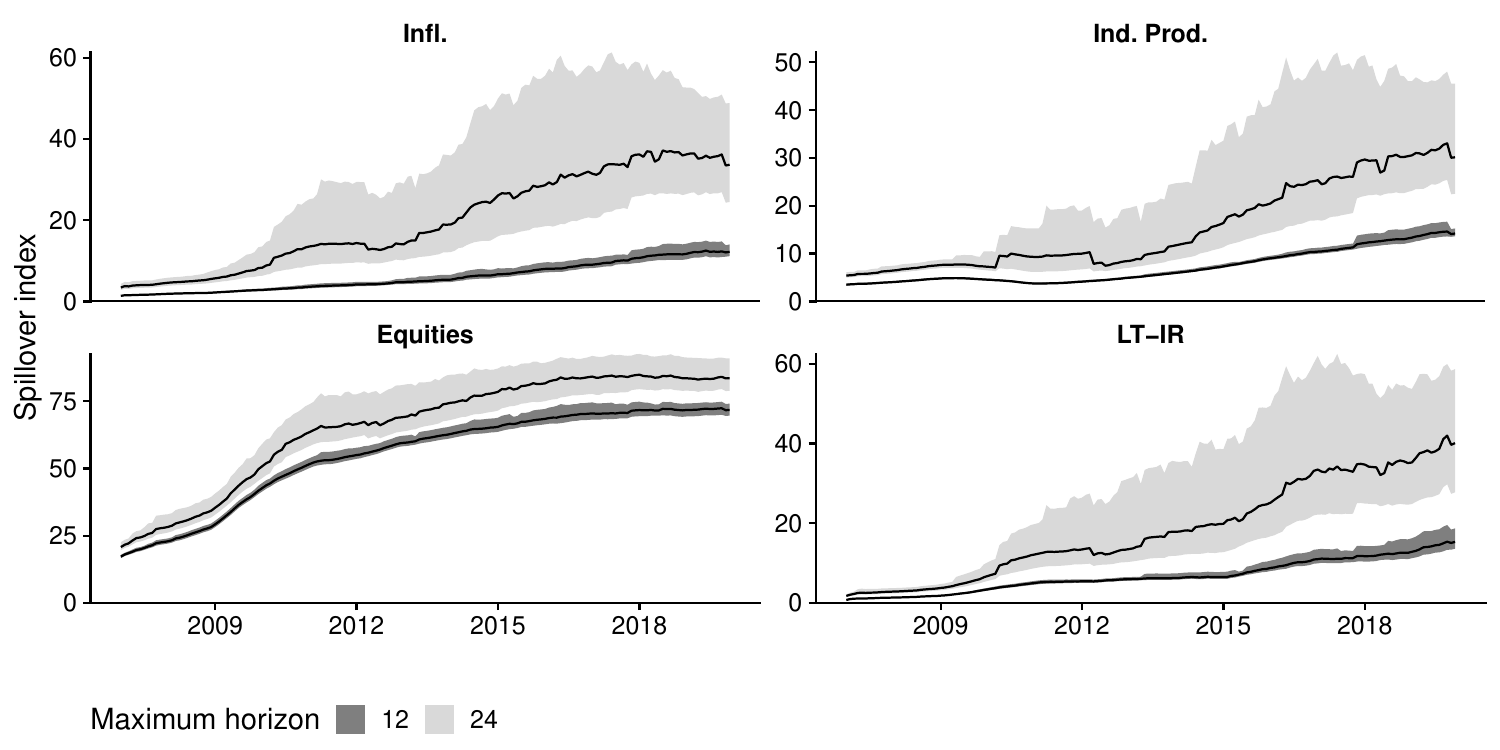}\\ 
    \caption{Diebold-Yilmaz spillover index by variable based on decomposition of forecast error variance of the full system.}\vspace*{-1em}
    \caption*{\footnotesize{\textit{Notes:} This index indicates the share of spillovers between countries one variable at a time (summed across countries). Estimated based on an expanding window. The solid line is the posterior median alongside the 68 percent posterior credible set.}}
     \label{DYindex_vars}
\end{figure}

The discussion up to this point focused on an overall measure of cross-country connectivity considering all focus variables jointly. To investigate whether connectivity plays a larger role for certain variables, we display the DY index for each of the focus variables separately in Figure \ref{DYindex_vars}. 
The figure reveals some interesting variable-specific differences. For example, the evolution of the DY index for industrial production, inflation and long-term interest rates is very similar to the behavior of the overall index displayed in Figure \ref{DYindex}. The degree of connectedness, is however, comparably smaller (35 percent). It is worth stressing that the posterior distribution is again strongly tilted towards higher levels of connectivity. A high degree of cross-country dependence between inflation rates is in line with \citet{Borio2007, Ciccarelli2010, km2018} who stress the importance of a global component in determining  domestic inflation rates.

Spillovers between equity prices show a distinct dynamic. The degree of connectivity in financial markets is generally higher compared to that of the remaining variables. The associated DY index is about 25 percent at the beginning of the hold-out period and rises sharply during the global financial crisis. This finding is in line with \cite{demirer2018estimating}, who demonstrate a strong increase in equity connectedness between banks during periods of financial stress. At the end of the sample period, the DY index amounts to about 75 percent. Notably, there is also little difference between the two forecast horizons. 

\begin{figure}
    \includegraphics[width=\textwidth]{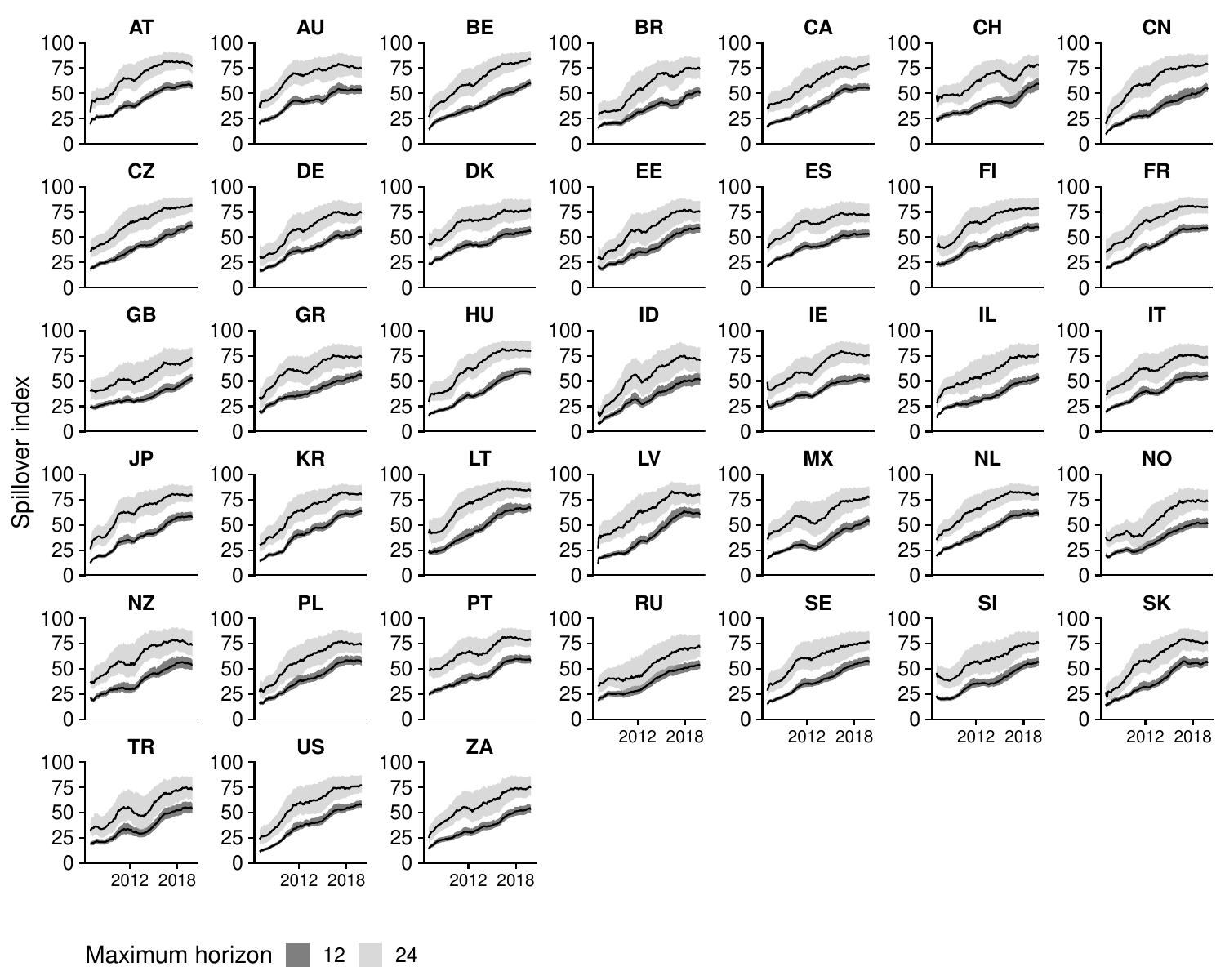}\\ 
    \caption{Diebold-Yilmaz spillover index by country based on decomposition of forecast error variance of the full system.}\vspace*{-1em}
    \caption*{\footnotesize{\textit{Notes:} This index indicates the share of foreign spillovers to the indicated country (summed across variables). Estimated based on an expanding window. The solid line is the posterior median alongside the 68 percent posterior credible set.}}
     \label{DYindex_cts}
\end{figure}

In Figure \ref{DYindex_cts}, we repeat the calculation and measure the role of cross-country effects for each country. In general, most of the countries considered display sustained increases in their respective DY indices peaking at about 75 percent. In some emerging economies, such as China and Indonesia, the index increases sharply right after 2008. 
Some notable exceptions are Turkey and Mexico which experienced a decrease in connectivity after the global financial crisis surrounding the Fed's tapering statement in June 2013. In the case of Greece, we observe a gradual decline in connectivity during the euro area sovereign debt crisis. This finding reflects a strong domestic component which complies with the fact that Greece was the epicenter of the crisis at the time. 

{Summing up and jointly considered with the forecasting results, the cross-variable DY indices paint a similar and consistent picture. Benefits from estimating a large multi-country system tend to increase with the forecast horizon. This finding is mirrored by increasing levels of connectivity if higher-order forecasts are considered. Our results indicate that the PVAR is capable of, in light of heavy shrinkage introduced through the Horseshoe prior, capturing cross-country relations flexibly. As opposed to other models (such as the FAVAR or the GVAR) the PVAR-IRGA introduces no particular restrictions on the coefficients. This feature is crucial to appropriately control for cross-country heterogeneity in light of a diverse set of countries such as the one we have in our data set.}

Appendix \ref{sec:resMN} provides additional results for the spillover indices computed using rolling windows of varying length. Using a rolling instead of an expanding window implies that past observations do not impact the estimates at some point. Thus, parameters are quicker to adjust to new information (at the cost of discarding past information). Results for both the overall and the variable-specific DY indices tell a similar story to the ones based on an expanding window. The main difference is that the estimates feature more movements in periods of economic turmoil such as the global financial crisis and the sovereign debt crisis in Europe.

\section{Concluding remarks}
Multi-country VARs have the potential to be enormous and simply working with unrestricted versions of them leads to over-parameterization and computational problems. The existing literature typically deals with these problems by imposing restrictions or reducing the dimension of the data. But the former strategy risks mis-specification and the latter risks losing information. Accordingly, in this paper, we have developed Bayesian methods for working with unrestricted VARs and rely on the Horseshoe prior to gain in parsimony by imposing shrinkage in a data-based fashion. Existing Bayesian work with VARs with such a prior has been done using MCMC methods. These are too computationally demanding to be used with PVARs with hundreds of dependent variables. In this paper, we have used IRGA methods to overcome this computational hurdle. We show that these allow for practical inference even in PVARs of huge dimension. Our macroeconomic empirical application demonstrates the benefits of being able to work with such large PVARs.     

\clearpage
\begin{appendix}
\begin{center}
    \textbf{\LARGE Appendix}
\end{center}
\setcounter{table}{0}
\setcounter{equation}{0}
\setcounter{figure}{0}
\renewcommand{\thetable}{A.\arabic{table}}
\renewcommand{\theequation}{A.\arabic{equation}}
\renewcommand{\thefigure}{A.\arabic{figure}}

\doublespacing\normalsize
\section{Vector approximate message passing}\label{sec:VAMP}
In this section, we provide a brief introduction on vector approximate message passing (VAMP). Before we provide details on the VAMP algorithm for the PVAR, a (very) brief introduction to message passing and the sum-product algorithm is in order.\footnote{For a textbook introduction, see \cite{bishop2006pattern}.} In general, message passing involves factorizing a joint density in an efficient way. Let $p(x_1, \dots, x_K)$ denote such a joint density over $K$ discrete random variables $\bm x = (x_1, \dots, x_K)'$. Moreover, let us assume that we are interested in the marginal distribution of some variable $x_j$ with $\bm x_{-j}$ denoting all  except  the $j^{th}$ variable. The marginal distribution of $x_j$ is then simply given by:
\begin{equation*}
p(x_j)  = \sum_{x_{-j}} p(\bm x)
\end{equation*}
which is a summation over $K-1$ random variables. For large $K$, calculating this marginal distribution using this formula becomes computationally infeasible. To get around this problem, we write the marginal density in a different way based on the concept of a factor graph. The factors are functions of the random variables. Under certain conditions (which are almost always fulfilled in econometric models) we can factorize the joint density $p(\bm x)$ as follows \citep[see, e.g., (8.59) in][]{bishop2006pattern}:
\begin{equation}
p(\bm x) = \prod_s f_s ( \bm x_s)
\end{equation}
with $\bm x_s$ denoting a subset of $\bm x$ and $f_s$ being a factor. These factors encode the relationship between variables.  We make use of this idea to represent the linear Bayesian regression model in terms of a factor graph and derive the relevant message passing algorithm which we then approximate using VAMP.

\begin{figure}
\begin{center}
\begin{subfigure}{0.4\linewidth}
    \caption{}
    \includegraphics[width=1\textwidth]{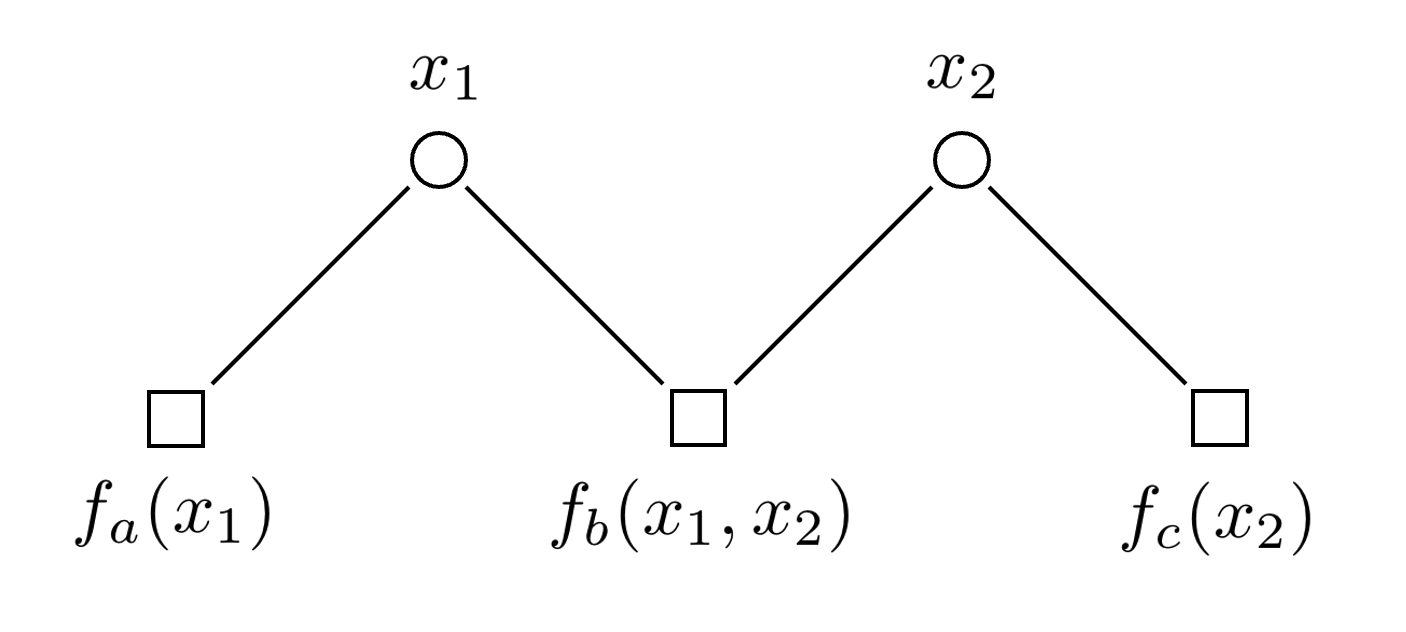}
\end{subfigure}
\begin{subfigure}{0.45\linewidth}
    \caption{}
    \vspace*{-0.4em}\includegraphics[width=1\textwidth]{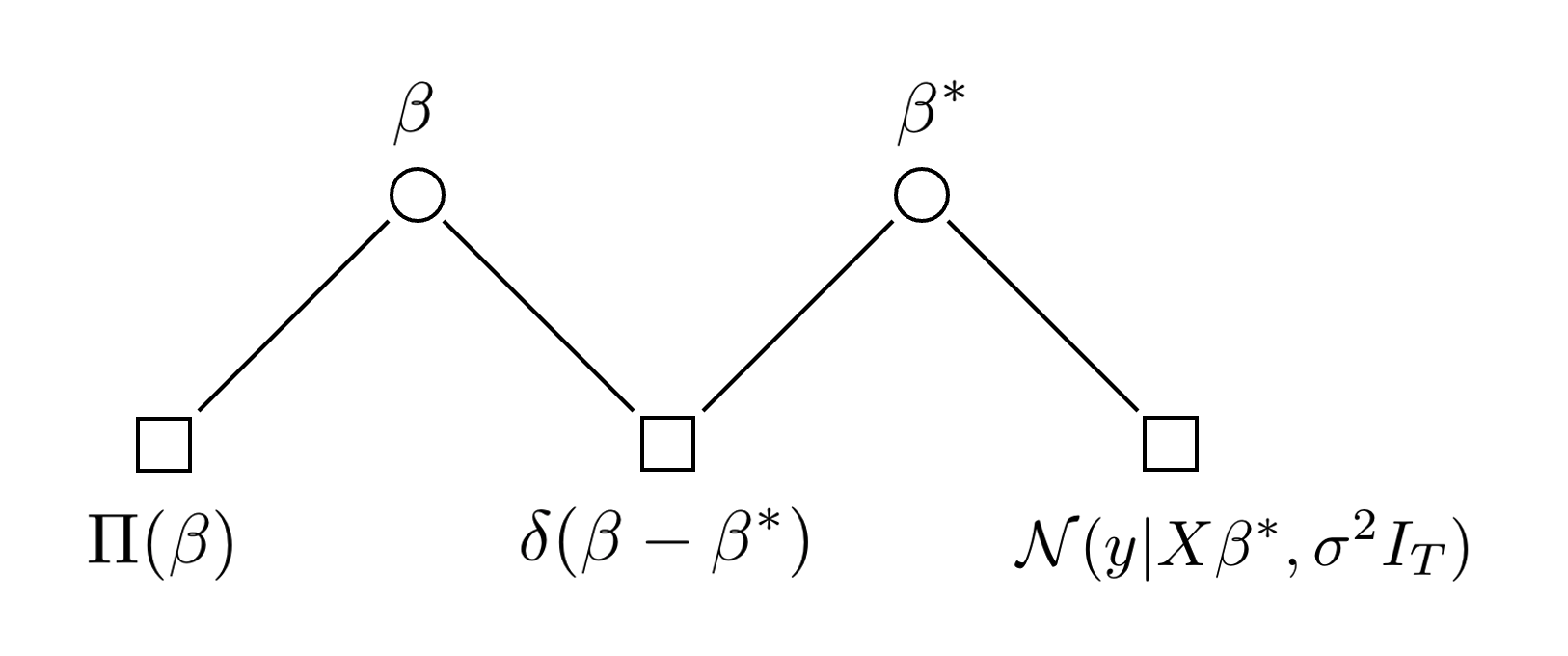}    
\end{subfigure}
\end{center}
\caption{Factor graphs.}
\label{fig:vamp}
\end{figure}

To set the stage,  let us assume that $\bm x = (x_1, x_2)'$ and these two random variables can be factorized as follows:
\begin{equation*}
p(x_1, x_2) = f_a(x_1) f_b (x_1, x_2)  f_c (x_2).
\end{equation*}
This implies that, for instance, the factor $f_b$ establishes a relationship between $x_1$ and $x_2$. The precise functional form of $f$ depends on the application. Notice that the factors in this decomposition depend on subsets of variables with $\bm x_a = \{x_1\}, \bm x_b = \{x_1, x_2\}$ and $\bm x_c = \{x_2\}$.  The corresponding factor graph is depicted in Figure \ref{fig:vamp}.

In this toy example the marginal distribution of, say, $x_1$ is proportional to:
\begin{align*}
p(x_1) &\propto \sum_{x_2}  p(x_1, x_2) \\
&= \sum_{x_2} f_a (x_1) f_b (x_1, x_2) f_c (x_2)\\
&= f_a(x_1) \sum_{x_2} f_b (x_1, x_2) f_c (x_2).
\end{align*}
Using the decomposition in the final line is computationally more efficient than computing the marginal distribution naively through the joint distribution. 

In what follows, we let $m_{f_i \to x_j}$ denote the message (or information) from factor node $f_i~(i \in \{a, b, c\})$  to variable $x_j~(j=1,2)$ whereas $m_{x_j \to f_i}$ refers to a message from a factor to a variable node. Since $f_c(x_3)$ depends exclusively on $x_3$ the message sent is equal to the factor node itself, $m_{f_c \to x_3} = f_c(x_3)$. These messages convey all relevant information from a factor to a node and vice versa. In the next step, we derive all relevant messages and show that the marginal distribution of $x_1$ is proportional to the product of all incoming messages.

Starting on the right part of Figure \ref{fig:vamp}(a), the message from factor node $f_c$ to $x_2$ is:
\begin{equation*}
m_{f_c \to x_2} = f_c(x_2).
\end{equation*}
The message from $x_2$ to factor node $f_b$ equals:
\begin{equation*}
m_{x_2 \to f_b} = f_c(x_2).
\end{equation*}
Finally, the messages from node $f_b$ to $x_1$ are:
\begin{equation*}
m_{f_b \to x_1} = \sum_{x_2} f_b(x_1, x_2)~  m_{x_2 \to f_b}.
\end{equation*}
Intuitively speaking, $m_{f_b \to x_1}$ captures the information flow from the nodes to the right of the variable node $x_1$. If we would use this information exclusively we would miss all information that arises from the node to the left of $x_1$. Because $f_a$ is an exterior node, the message from $f_a$ to $x_1$ is simply $m_{f_a \to x_1} = f_a(x_1)$.  

The product over all incoming messages is proportional to the marginal distribution $p(x_1)$:
\begin{equation}
p(x_1) \propto m_{f_b \to x_1} ~  m_{f_a \to x_1} = m_{f_b \to x_1} ~ f_a(x_1). \label{eq: x1}
\end{equation}
Deriving the marginal distribution of $x_2$ analogously yields:
\begin{equation}
p(x_2) \propto m_{f_b \to x_2} \mu_{f_c \to x_2} = \mu_{f_b \to x_2} ~f_c(x_2). \label{eq: x2}
\end{equation}
\cite{bishop2006pattern} highlights that there exists a close relationship between the belief about a variable $x_j$ (which is defined as the product of all incoming messages to the variable node), $b(x_j)$, and the corresponding marginal distribution. If the graphical model is a tree, the beliefs converge to the marginal distribution after one iteration of a message passing algorithm. 

Algorithms exploiting Eqs. (\ref{eq: x1}) and (\ref{eq: x2}) are labeled sum-product algorithms \citep[see, e.g.,][]{korobilis2019high}. Notice that to arrive at the marginals of $x_1$ ($x_2$), we need to marginalize over $x_2$ ($x_1$). This summation is often difficult to compute. As a solution, researchers often rely on approximations to these sums (or integrals more generally) and arrive at so-called approximate message passing algorithms \citep{donoho2009message}. 

Before we discuss such approximations in more detail, we show how to design a message passing algorithm for the general regression model. Let $\bm y$ denote a $T-$ dimensional response vector and $\bm X$ is a $T \times K$ matrix of regressors. The corresponding regression coefficients are denoted by $\bm \beta$ and the error variance is given by $\sigma^2$. Furthermore, we let $\Pi(\bm \beta)$ denote the prior. We assume throughout that any hyperparameters associated with the prior and $\sigma^2$ are known.\footnote{In practice, we will update them using expectation maximization steps using the quantities presented below.}

The posterior of $\bm \beta$ (conditional on $\sigma^2$) is given by:
\begin{equation}
p(\bm \beta | \sigma^2, \bm y, \bm X) \propto \mathcal{N}(\bm y| \bm X \bm \beta, \sigma^2 \bm I_T) ~ \Pi(\bm \beta).\label{eq: posteriorB}
\end{equation}
To derive the graphical representation of the regression model it proves convenient to introduce a copy of $\bm \beta$, labeled $\bm \beta^*$. This allows us to rewrite the posterior in (\ref{eq: posteriorB}) as follows:
\begin{equation}
p(\bm \beta | \sigma^2, \bm y, \bm X) \propto \mathcal{N}(\bm y| \bm X \bm \beta^*, \sigma^2 \bm I_T) \delta(\bm \beta - \bm \beta^*) \Pi (\bm \beta), \label{eq: posteriorBstar} 
\end{equation}
with $\delta$ denoting the Dirac Delta function which equals one if $\bm \beta = \bm \beta^*$.

The factorization in (\ref{eq: posteriorBstar}) closely resembles the one used in our toy example. If $\bm x_1 = \bm \beta$, $f_a(\bm x_1) = \Pi(\bm \beta)$, $\bm x_2 = \bm \beta^*$, $f_b(\bm x_1, \bm x_2) = \delta (\bm \beta - \bm \beta^*)$ and $f_c(\bm x_2) = \mathcal{N}(\bm y| \bm X \bm \beta^*, \sigma^2 \bm I_T)$ we can easily derive the belief of $\bm \beta$ and $\bm \beta^*$ as follows:
\begin{align}
b(\bm \beta) &\propto  m_{\Pi \to \bm \beta} ~m_{\delta \to \bm \beta} = m_{\delta \to \bm \beta}~ \Pi(\bm \beta) \label{eq: beta_mp}\\
b(\bm \beta^*)  &\propto m_{\delta \to \bm \beta^*} ~ m_{\mathcal{N} \to \bm \beta^*} = \mathcal{N}(\bm y | \bm x \bm \beta^*, \sigma^2 \bm I_T) ~ m_{\delta \to \bm \beta^*}. \label{eq: betastar_mp}
\end{align}
\cite{irga} show how (\ref{eq: beta_mp}) and (\ref{eq: betastar_mp}) can be used to derive a message passing algorithm which exploits certain properties of the structure of the graphical model outlined above. This message  passing algorithm cycles between the following updating steps:
\begin{align}
b(\bm \beta) &\propto m_{\delta \to \bm \beta} ~ \Pi(\bm \beta), \label{eq: belief_beta}\\
m_{\beta \to \delta} &\propto \frac{b(\bm \beta)}{m_{\delta \to \bm \beta}}, \\
b(\bm \beta^*) &\propto m_{\bm \beta \to \delta}~\mathcal{N}(\bm y | \bm X \bm \beta^*, \sigma^2 \bm I_T), \\
m_{\delta \to \bm \beta^*} = m_{\bm \beta^* \to \delta} &\propto \frac{b(\bm \beta^*)}{m_{\bm \beta \to \delta}}.
\end{align}
The corresponding updates are then obtained as follows.
Initialize $m_{\delta \to \beta} = \mathcal{N}(\bm \beta| \bm \xi^{(0)}, \zeta^{(0)} \bm I_K)$. For iteration $j$, the belief about $\bm \beta$ is approximated through a Gaussian distribution
\begin{equation}
 b(\bm \beta) \approx  \mathcal{N}( \bm \beta | \overline{\bm \beta}^{(j)}, s^{(j)} \bm I_K),
\end{equation}
with $\overline{\bm \beta}^{(j)} = \mathbb{E}_{b(\bm \beta)}(\bm \beta)$ and $s^{(j)} = \text{Tr}(\bm \Omega^k)/K$ with $\bm \Omega^{(j)} = \text{Cov}_{b(\bm \beta)}(\bm \beta)$ with tr denoting the trace operator. Using (\ref{eq: belief_beta}) yields the first updating step of the algorithm:
\begin{equation}
\overline{\beta}^j_i = \mathbb{E}(\beta_i | \xi^{(j)}_i, \zeta^{(j)}) \text{ with } s^{(j)} = \sum_{i=1}^K \text{Var}(\beta_i| \xi^{(j)}_i, s^j). \label{eq:first}
\end{equation}
Given that we approximate $m_{\delta \to \beta}$ using a Gaussian distribution and, under a Gaussian prior, it directly follows that $b(\bm \beta)$ is Gaussian as well, the message $m_{\bm \beta \to \delta}$ is also Gaussian with:
\begin{equation}
m_{\bm \beta \to \delta} \sim \mathcal{N}(\bm \beta^* | \bm \xi^{*j}, \zeta^{*j} \bm I_K ),
\end{equation}
where
\begin{align*}
1/\zeta^{* {(j)}} = 1/s^{(j)} - 1/\zeta^{(j)}, \\
\xi^{* (j)} = \frac{\zeta^{(j)} \overline{\bm \beta}^{(j)} - s^{(j)} \bm \xi}{\zeta^{(j)} - s^{(j)}}.
\end{align*}
In the next step, we notice that because $m_{\bm \beta \to \delta}$ is Gaussian and the likelihood based on the copy of $\bm \beta$ is Gaussian too, the  belief $b(\bm \beta^*)$ is also approximated using a Gaussian distribution with a covariance matrix that is proportional to the identity matrix \citep{rangan2019vector,irga}. The corresponding approximating density is:
\begin{equation}
b(\bm \beta^*) \approx \mathcal{N}(\bm \beta^* | \overline{\bm \beta}^{* (j)}, s^{* (j)} \bm I_K).
\end{equation}
Simply computing the moments of this distribution requires inverting a $K\times K$ matrix. However, the formula for these moments can be written in a different way so as to avoid inverting  such a matrix by decomposing $\bm X$ using the singular value decomposition (SVD) of $\bm X = \bm U \bm D \bm V'$. The corresponding mean and variance are then given by:
\begin{align*}
\overline{\bm \beta}^{* (j)} &= \xi^{* (j)} + \bm V (\sigma^2 \bm D^{-1}/\zeta^{* {(j)}} + \bm D)^{-1} (\bm U' \bm y - \bm D \bm V' \xi^{* (j)}), \\
s^{* (j)} &= \zeta^{* (j)} (1 - \text{Tr}(\bm V (\sigma^2 \bm D^{-1}/\zeta^{* {(j)}} + \bm D)^{-1})/K).
\end{align*}
These expressions require only the inversion of diagonal matrices and are thus computationally cheap to implement. 

Finally, we set:
\begin{align}
1/\zeta^{{(j+1)}} = 1/s^{*(j)} - 1/\zeta^{*(j)}, \\
\xi^{(j+1)} = \frac{\zeta^{*(j)} \overline{\bm \beta}^{*(j)} - s^{*(j)} \bm \xi}{\zeta^{*(j)} - s^{*(j)}}. \label{eq: finalstep}
\end{align}
Repeating steps (\ref{eq:first}) to (\ref{eq: finalstep}) $S$ times yields the approximate posterior which is $\mathcal{N}(\overline{\bm \beta}^{(S)}, s^{(S)} \bm I_K)$. In practice, we stop iterating the algorithm if $||\overline{\bm \beta}^{(j)} - \overline{\bm \beta}^{(j-1)}||^2$ is smaller than a certain threshold close to zero. This algorithm shows good convergence properties, is computationally fast and has been shown to work well empirically. 

\subsection{Updating the error variance}
Conditional on the data and the current state of the parameter vector $\bm{\beta}$, let $\bm{\epsilon} = \bm{y} - \bm{X}\bm{\beta}$. Under the prior $\sigma^2\sim\mathcal{G}^{-1}(a_\sigma,b_\sigma)$, the error variance update is given by:
\begin{equation*}
    \sigma^2 = 2 b_\sigma + \frac{\bm{\epsilon}'\bm{\epsilon}}{2 a_\sigma + T}.
\end{equation*}

\subsection{Updating the Horseshoe prior}
Based on the general Horseshoe prior described in Sub-Section \ref{sec:priors}, the updating quantities for the hierarchical prior structure in the VAMP algorithm for a parameter vector $\bm{\phi}=(\phi_1,\hdots,\phi_k)'$ are given by:
\begin{align*}
    \psi_i^{-2} &= \frac{1}{\nu_i^{-1} + \phi_i^2\lambda^{-2}/2},\quad
    \lambda^{-2} = \frac{k+1}{2\xi^{-1} + \psi_i^{-2}\sum_{i=1}^{k}\phi_i^2}\\
    \nu_i &= \frac{1}{1+\psi_i^{-2}}, \quad \xi = \frac{1}{1+\lambda^{-2}}.
\end{align*}
Note that in the empirical application, we have country- and equation-specific global and local shrinkage parameters.

\setcounter{table}{0}
\setcounter{equation}{0}
\setcounter{figure}{0}
\renewcommand{\thetable}{B.\arabic{table}}
\renewcommand{\theequation}{B.\arabic{equation}}
\renewcommand{\thefigure}{B.\arabic{figure}}

\section{Inference for PVAR-IRGA} \label{sec: posterior}
\subsection{MCMC sampling of the VAR coefficients}
Let $\tilde{\bm y}_{ij} = \bm Q'_{i1}(\bm y_{ij} - \tilde{\bm z}_{ij} \overline{\bm B}_{ij, \bullet})$, $\bm \tilde{\bm x}_{ij} = (\tilde{\bm Q}'_{i1} \bm x_i)$ and $\bm{\Sigma}_{ij} = \bm Q'_{i1} \tilde{\bm z}_{ij} \overline{\bm V}_{ij, \bullet} \tilde{\bm z}'_{ij} \bm Q_{i1} + \sigma^2_{\varepsilon, ij} \bm I_{k}$, and recall that the likelihood of the approximate model is:
\begin{equation*}
    \bm Q'_{i1}(\bm y_{ij} - \tilde{\bm z}_{ij} \overline{\bm B}_{ij, \bullet}) \sim \mathcal{N}(\bm Q_{i1}' \bm x_{i} \bm A_{ij, \bullet}, \bm Q'_{i1} \tilde{\bm z}_{ij} \overline{\bm V}_{ij, \bullet} \tilde{\bm z}'_{ij} \bm Q_{i1} + \sigma^2_{\varepsilon, ij} \bm I_{k}).
\end{equation*}
The conditional posterior distribution of the VAR coefficients collected in $\bm A_{ij, \bullet}$ is:
\begin{align*}
    \bm A_{ij, \bullet} &\sim \mathcal{N}(\bm{\bar\mu}_{ij},\bm{\bar S}_{ij}),\\
    \bm{\bar S}_{ij} &= (\bm{V}_{A,ij}^{-1} + \bm \tilde{\bm x}_{ij}'\bm{\Sigma}_{ij}^{-1}\tilde{\bm x}_{ij})^{-1},\\
    \bm{\bar\mu}_{ij} &= \bm{\bar S}_{ij}(\bm{V}_{A,ij}^{-1}\bm{\mu}_{A,ij} + \bm \tilde{\bm x}_{ij}'\bm{\Sigma}_{ij}^{-1}\tilde{\bm y}_{ij}),\\
\end{align*}
where, $\bm{\mu}_{A,ij}$ is the prior mean and $\bm{V}_{A,ij}$ the prior covariance matrix. For our empirical application, the prior mean is set to a zero-vector, $\bm{\mu}_{A,ij} = \bm{0}_k$ for all equations and countries, and the prior variance is defined by the Horseshoe prior: $\bm{V}_{A,ij} = (\lambda_{A,ij})^2 \times \text{diag}(\psi_{ij,1}^2,\hdots,\psi_{ij,k}^2)$.

\subsection{Minnesota prior for PVAR-IRGA}\label{sec:minnesota}
The Minnesota prior for PVAR-IRGA, in the spirit of \citet{chan2021minnesota}, is constructed as follows. The global scaling parameter for equation $i$, $\lambda_{i}\sim\mathcal{C}^{+}(0,1)$, follows a half-Cauchy distribution as in the case of the Horseshoe. This parameter governs the overall tightness of the Minnesota prior on an equation-by-equation basis. Let $\hat{\sigma}_{i}^2$ denote the residual variance of a univariate AR($p$) model for the $i$th variable. The local scales $\psi_{ij}$ are set deterministically in the Minnesota tradition, by introducing a quadratic lag-penalty scaled by the ratio $\hat{\sigma}_{i}^2/\hat{\sigma}_{j}^2$ for the shrinkage parameter related to variable $j$.

\subsection{Posteriors related to the Horseshoe prior}
We present the conditional posterior distributions of the Horseshoe prior for the general example in Sub-Section \ref{sec:priors} to economize on notation. Note that for the PVAR, we have country- and equation-specific global and local shrinkage parameters. 

Our implementation is based on the auxiliary representation of the Horseshoe prior discussed in \citet{MS2016}. For a $k$-dimensional parameter vector $\bm{\phi} = (\phi_1,\hdots,\phi_k)'$, with $\phi_j$ indexing the $j=1,\hdots,k$ coefficient, we obtain the following inverse Gamma distributed posteriors:
\begin{align*}
    \psi_j^2|\phi_j,\lambda,\nu_j &\sim \mathcal{G}^{-1}\left(1,\frac{1}{\nu_j} + \frac{\phi_j^2}{2\lambda^2}\right), \quad \lambda^2|\phi_j,\psi_j,\xi\sim\mathcal{G}^{-1}\left(\frac{k+1}{2}, \frac{1}{\xi} + \sum_{j=1}^k\frac{\phi_j^2}{2\psi_j^2}\right),\\
    \nu_j|\psi_j&\sim\mathcal{G}^{-1}\left(1,1+\frac{1}{\psi_j^2}\right),\quad 
    \xi|\lambda\sim\mathcal{G}^{-1}\left(1,1+\frac{1}{\lambda^2}\right).
\end{align*}
For the case of the Minnesota prior, only the auxiliary variables related to the global parameter need to be drawn.

\setcounter{table}{0}
\setcounter{equation}{0}
\setcounter{figure}{0}
\renewcommand{\thetable}{C.\arabic{table}}
\renewcommand{\theequation}{C.\arabic{equation}}
\renewcommand{\thefigure}{C.\arabic{figure}}

\section{Other models}\label{app:othermodels}
\subsection{Single-country vector autoregressions}
The Bayesian vector autoregressions (BVARs) for single countries are given by:
\begin{equation*}
    \bm{y}_{it} = \bm{\Gamma}_{i1} \bm{y}_{it-1} + \hdots + \bm{\Gamma}_{ip} \bm{y}_{it-p} + \bm{\epsilon}_{it}, \quad \bm{\epsilon}_{it}\sim\mathcal{N}(\bm{0},\bm{\Sigma}_i).
\end{equation*}
Note that this specification corresponds to Eq. (\ref{eq:likelihood}) with $\bm{\Xi}_i = \bm{0}_{M\times K_{other}}$, ruling out both dynamic and static interdependencies across countries by assuming $Cov(\bm{\epsilon}_{it},\bm{\epsilon}_{st}) = \bm{0}$. The models are estimated one country at a time. 

We use a Horseshoe prior on the own and cross-variable VAR coefficients. Alternatively, we use a hierarchical Minnesota-type prior similar to the one discussed in \ref{sec:minnesota}. The VARs are estimated equation-by-equation. The priors on the covariances are independent Gaussian distributions with mean zero and variance $10$. The variances of the structural errors are assigned independent weakly informative inverse Gamma priors.

\subsection{Factor-augmented vector autoregressions}
Define a vector $\tilde{\bm{y}}_{it} = (\bm{y}_{it}',\bm{f}_{it}')'$, with $\bm{y}_{it}$ including the domestic variables of country $i$ and $\bm{f}_{it}$ denoting a set of principal components extracted from the matrix $\bm{F}_i = (\bm{y}_{-i,1},\hdots,\bm{y}_{-i,T})'$. The domestic variables and the ``foreign'' factors $\bm{f}_{it}$ are modeled jointly in a VAR: 
\begin{equation*}
    \tilde{\bm{y}}_{it} = \bm{\Gamma}_{i1} \tilde{\bm{y}}_{it-1} + \hdots + \bm{\Gamma}_{ip} \tilde{\bm{y}}_{it-p} + \tilde{\bm{\epsilon}}_{it}, \quad \tilde{\bm{\epsilon}}_{it}\sim\mathcal{N}(\bm{0},{\bm{\Sigma}}_i).
\end{equation*}
We use the same hierarchical Minnesota and Horseshoe priors and estimation algorithm as for the BVAR and estimate the models one country at a time.

\subsection{Global vector autoregression}
The global vector autoregressive model (GVAR) incorporates cross-country dependencies via including weighted averages of the foreign variables in the domestic equations. In particular, the GVAR is given by: 
\begin{equation*}
    \bm{y}_{it} = \bm{\Gamma}_{i1} \bm{y}_{it-1} + \hdots + \bm{\Gamma}_{ip} \bm{y}_{it-p} + \bm{\Xi}_{i0} \bm{y}_{it}^{\ast} + \bm{\Xi}_{i1} \bm{y}_{it-1}^{\ast} + \hdots + \bm{\Xi}_{iq} \bm{y}_{it-q}^{\ast} + \bm{\epsilon}_{it}, \quad \bm{\epsilon}_{it}\sim\mathcal{N}(\bm{0},\bm{\Sigma}_i),
\end{equation*}
where the $\bm{y}_{it}^{\ast}$ are defined as
\begin{equation*}
    \bm{y}_{it}^{\ast} = \sum_{j=1}^N w_{ij} \bm{y}_{jt}.
\end{equation*}
The pre-specified set of weights staisfy the restrictions $\sum_{j=1}^N w_{ij} = 1$ and $w_{ii} = 0$. We estimate the GVAR using the \texttt{R}-package by \citet{bock2020bgvar}, using again a hierarchical Minnesota and a Horseshoe prior similar to the BVAR and the FAVAR.

\setcounter{table}{0}
\setcounter{equation}{0}
\setcounter{figure}{0}
\renewcommand{\thetable}{D.\arabic{table}}
\renewcommand{\theequation}{D.\arabic{equation}}
\renewcommand{\thefigure}{D.\arabic{figure}}

\section{Additional empirical results}\label{sec:resMN}
\subsection{Minnesota prior}
To provide empirical evidence on the role of the Horseshoe prior that we choose to impose shrinkage in the baseline specification of PVAR-IRGA, we re-do the forecast exercise with all models estimated using a Minnesota-type prior. Table \ref{tb3} shows the forecast results for this set of models relative to PVAR-IRGA estimated with a Minnesota prior. Table \ref{tb4} provides a direct comparison between the two competing priors for PVAR-IRGA. On average, the performance of PVAR-IRGA-HS is stronger than PVAR-IRGA-MN. However, it is worth mentioning that PVAR-IRGA-MN produces modest improvements in our aggregate measures of predictive accuracy for equity prices and long-term interest rates with respect to both point and density forecasts.

\begin{table}[!ht]
\scriptsize
\caption{Summary of Forecast Exercise, all models with Minnesota prior.}\label{tb3}\vspace*{-1.5em}
\begin{center}
\begin{tabular}{llcrrrrcrrrr}
\toprule
\multicolumn{1}{l}{\bfseries }&\multicolumn{1}{c}{\bfseries }&\multicolumn{1}{c}{\bfseries }&\multicolumn{4}{c}{\bfseries MAE}&\multicolumn{1}{c}{\bfseries }&\multicolumn{4}{c}{\bfseries LPS}\tabularnewline
\cline{4-7} \cline{9-12}
\multicolumn{1}{l}{}&\multicolumn{1}{c}{Model}&\multicolumn{1}{c}{}&\multicolumn{1}{c}{Equities}&\multicolumn{1}{c}{Ind. prod.}&\multicolumn{1}{c}{LT-IR}&\multicolumn{1}{c}{Infl.}&\multicolumn{1}{c}{}&\multicolumn{1}{c}{Equities}&\multicolumn{1}{c}{Ind. prod.}&\multicolumn{1}{c}{LT-IR}&\multicolumn{1}{c}{Infl.}\tabularnewline
\midrule
{\itshape h=1}&&&&&&&&&&&\tabularnewline
\shadeRow   ~~&   BVAR&   &   1.003&   0.992&   0.955&   0.965&   &     -5.877&      0.058&     6.796&      3.241\tabularnewline
\shadeRow   ~~&   &   &   (42.1)&   (16.1)&   (40.0)&   (45.7)&   &   ( 7.9)&   (29.0)&   (40.0)&   (31.4)\tabularnewline
   ~~&   FAVAR-10&   &   1.003&   1.007&   0.948&   0.971&   &     -4.905&     -0.789&     8.324&      3.417\tabularnewline
   ~~&   &   &   (31.6)&   (12.9)&   (56.7)&   (45.7)&   &   (15.8)&   (22.6)&   (60.0)&   (40.0)\tabularnewline
\shadeBench   ~~&   PVAR-IRGA&   &   0.829&   0.414&   0.265&   0.397&   &   -178.797&    -42.377&    43.308&    -18.394\tabularnewline
\shadeBench   ~~&   &   &   (23.7)&   (54.8)&   ( 3.3)&   ( 8.6)&   &   (63.2)&   (38.7)&   ( 0.0)&   (28.6)\tabularnewline
   ~~&   GVAR&   &   1.045&   1.083&   1.127&   1.166&   &    -11.543&    -18.524&   -49.750&    -39.472\tabularnewline
   ~~&   &   &   ( 2.6)&   (16.1)&   ( 0.0)&   ( 0.0)&   &   (13.2)&   ( 9.7)&   ( 0.0)&   ( 0.0)\tabularnewline
\midrule
{\itshape h=12}&&&&&&&&&&&\tabularnewline
\shadeRow   ~~&   BVAR&   &   1.000&   1.033&   0.940&   1.010&   &     -1.092&    -11.992&     5.480&      8.696\tabularnewline
\shadeRow   ~~&   &   &   (42.1)&   ( 6.5)&   (40.0)&   (28.6)&   &   (15.8)&   (16.1)&   (26.7)&   (45.7)\tabularnewline
   ~~&   FAVAR-10&   &   0.997&   1.034&   0.923&   1.008&   &     -0.220&    -16.291&     9.441&      5.539\tabularnewline
   ~~&   &   &   (31.6)&   ( 9.7)&   (43.3)&   (20.0)&   &   (10.5)&   (12.9)&   (56.7)&   (17.1)\tabularnewline
\shadeBench   ~~&   PVAR-IRGA&   &   0.856&   0.700&   0.548&   0.772&   &   -192.860&   -158.364&   -95.594&   -184.650\tabularnewline
\shadeBench   ~~&   &   &   (26.3)&   (45.2)&   (16.7)&   (42.9)&   &   (68.4)&   (48.4)&   (16.7)&   (31.4)\tabularnewline
   ~~&   GVAR&   &   1.071&   1.018&   1.021&   1.115&   &    -13.015&     -4.675&   -25.273&    -20.469\tabularnewline
   ~~&   &   &   ( 0.0)&   (38.7)&   ( 0.0)&   ( 8.6)&   &   ( 5.3)&   (22.6)&   ( 0.0)&   ( 5.7)\tabularnewline
\bottomrule
\end{tabular}
\end{center}
\vspace*{-1em}\tiny{\textit{Notes:} GDP-weighted average over countries, win percentage across countries in parentheses. Root mean squared error (RMSE) and log predictive score (LPS) relative to the benchmark. The benchmark PVAR-IRGA (shaded in red) shows actual values, all other models are in ratios to the benchmark for RMSEs and in differences for LPSs.}
\end{table}

\begin{table}[!ht]
\scriptsize
\caption{Summary of Forecast Exercise, comparison of PVAR-IRGA estimated with Horseshoe (HS) and Minnesota (MN) prior.}\label{tb4}\vspace*{-1.5em}
\begin{center}
\begin{tabular}{llcrrrrcrrrr}
\toprule
\multicolumn{1}{l}{\bfseries }&\multicolumn{1}{c}{\bfseries }&\multicolumn{1}{c}{\bfseries }&\multicolumn{4}{c}{\bfseries MAE}&\multicolumn{1}{c}{\bfseries }&\multicolumn{4}{c}{\bfseries LPS}\tabularnewline
\cline{4-7} \cline{9-12}
\multicolumn{1}{l}{}&\multicolumn{1}{c}{Model}&\multicolumn{1}{c}{}&\multicolumn{1}{c}{Equities}&\multicolumn{1}{c}{Ind. prod.}&\multicolumn{1}{c}{LT-IR}&\multicolumn{1}{c}{Infl.}&\multicolumn{1}{c}{}&\multicolumn{1}{c}{Equities}&\multicolumn{1}{c}{Ind. prod.}&\multicolumn{1}{c}{LT-IR}&\multicolumn{1}{c}{Infl.}\tabularnewline
\midrule
{\itshape h=1}&&&&&&&&&&&\tabularnewline
\shadeBench   ~~&   PVAR-IRGA-HS&   &   0.825&   0.410&   0.274&   0.386&   &   -177.799&    -39.815&     36.181&    -13.387\tabularnewline
\shadeBench   ~~&   &   &   (63.2)&   (51.6)&   (46.7)&   (77.1)&   &   (39.5)&   (61.3)&   (46.7)&   (77.1)\tabularnewline
   ~~&   PVAR-IRGA-MN&   &   1.005&   1.009&   0.967&   1.029&   &     -0.998&     -2.562&      7.127&     -5.007\tabularnewline
   ~~&   &   &   (36.8)&   (48.4)&   (53.3)&   (22.9)&   &   (60.5)&   (38.7)&   (53.3)&   (22.9)\tabularnewline
\midrule
{\itshape h=12}&&&&&&&&&&&\tabularnewline
\shadeBench   ~~&   PVAR-IRGA-HS&   &   0.875&   0.664&   0.570&   0.759&   &   -204.541&   -149.824&   -104.962&   -174.713\tabularnewline
\shadeBench   ~~&   &   &   (47.4)&   (61.3)&   (63.3)&   (65.7)&   &   (36.8)&   (77.4)&   (43.3)&   (74.3)\tabularnewline
   ~~&   PVAR-IRGA-MN&   &   0.979&   1.054&   0.962&   1.017&   &     11.681&     -8.540&      9.368&     -9.938\tabularnewline
   ~~&   &   &   (52.6)&   (38.7)&   (36.7)&   (34.3)&   &   (63.2)&   (22.6)&   (56.7)&   (25.7)\tabularnewline
\bottomrule
\end{tabular}
\end{center}
\vspace*{-1em}\tiny{\textit{Notes:} GDP-weighted average over countries, win percentage across countries in parentheses. Root mean squared error (RMSE) and log predictive score (LPS) relative to the benchmark. The benchmark PVAR-IRGA-HS (shaded in red) shows actual values, PVAR-IRGA-MN is in ratios to the benchmark for RMSEs and in differences for LPSs.}
\end{table}

\subsection{Ordering of variables}\label{sec: rob_ordering}
Our implementation of the equation-by-equation estimation algorithm of the PVAR implies that the posterior estimates are not invariant with respect to different orderings of the variables. 

To shed light on the robustness of our approach in this context, we assess the sensitivity of the posterior median estimates of the spillover index to different orderings of country blocks (defined by the four categories Advanced European, Emerging European, Advanced Other and Emerging Other) in the system. We choose this approach for two reasons. First, it is not possible to compute all possible permutations of individual variables. Consequently, we compute the same metric, the spillover index, for all possible permutations of the four country-blocks. Second, it is infeasible to show robustness with respect to estimates of all parameters of the model. Thus, we rely on the spillover index as a relevant summary statistic for this purpose. 

\begin{figure}[!ht]
    \centering
    \includegraphics[width=\textwidth]{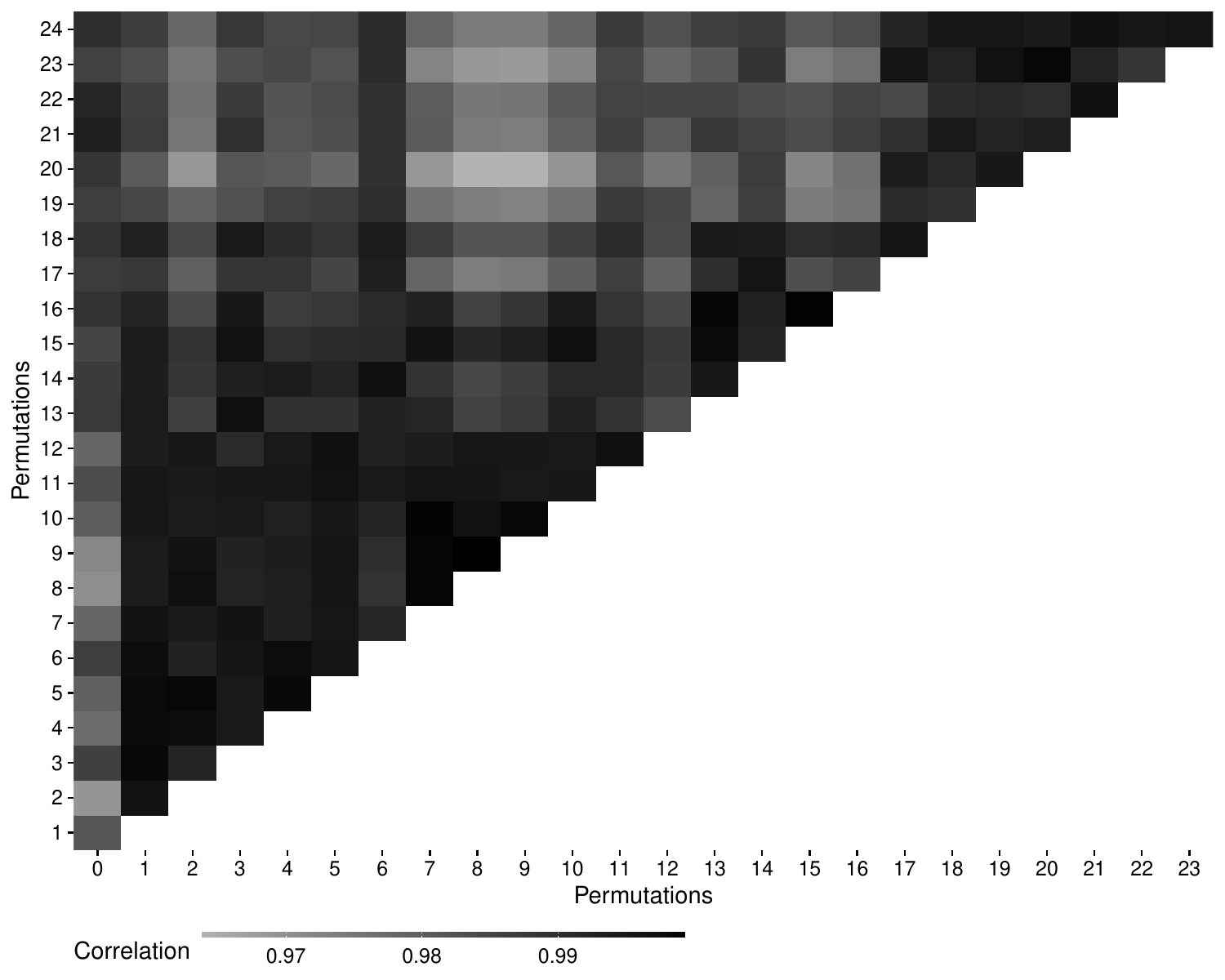}
    \caption{Correlation of spillover index estimates across permutations of country blocks.}
    \label{fig:ordering-robustness}
\end{figure}

Figure \ref{fig:ordering-robustness} shows a pairwise correlation matrix of the posterior estimates of the spillover index across all possible permutations of country blocks. The correlations range from $0.97$ to close to one, suggesting almost identical estimates of the spillover indices across orderings. Hence, we conclude that the ordering of the variables in our model  only has negligible effects on the results.

\subsection{Spillover index with rolling window}\label{sec: rob_rollingwindow}
\begin{figure}[!ht]
    \includegraphics[width=\textwidth]{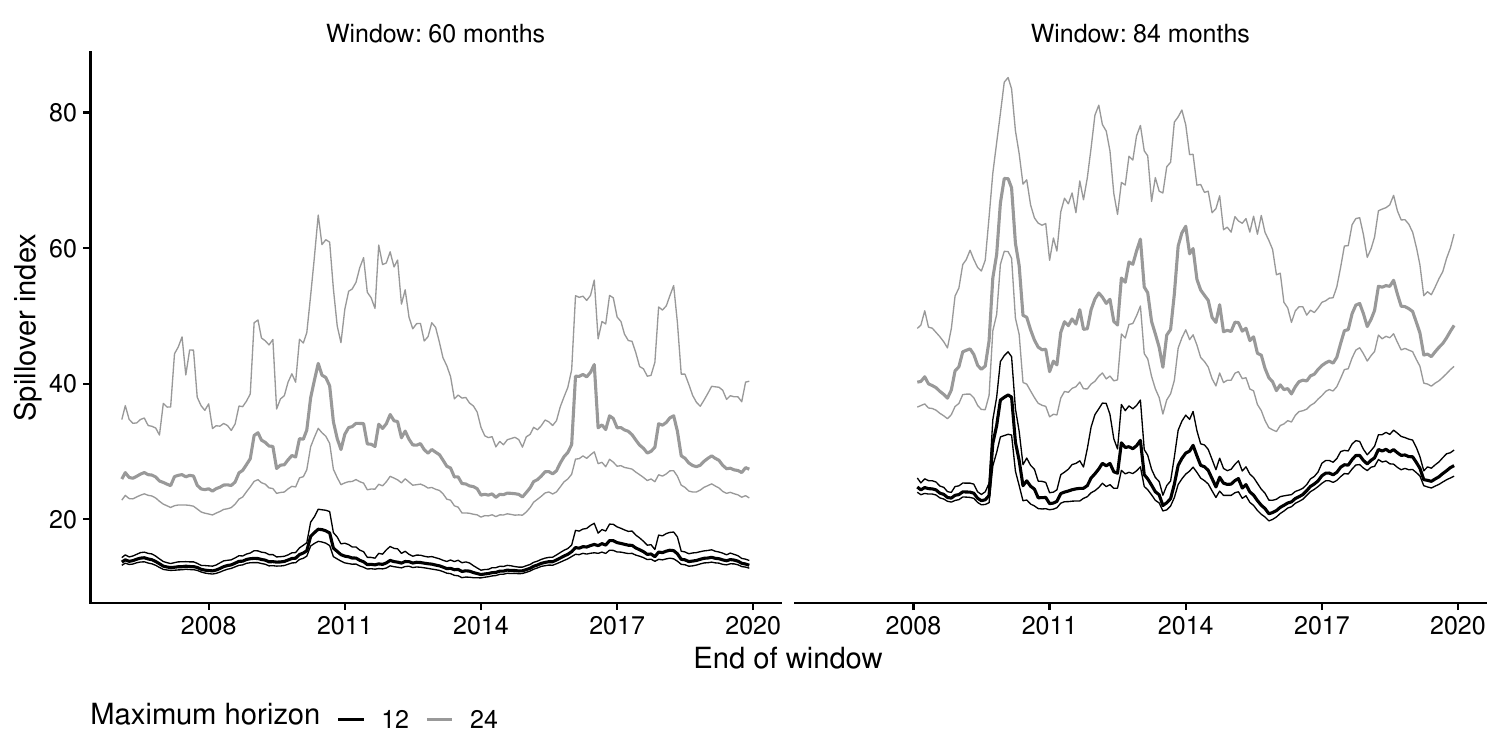}\\ 
    \caption{Diebold-Yilmaz spillover index for all countries based on decomposition of forecast error variance of the full system using a rolling window of observations.}\vspace*{-1em}
    \caption*{\footnotesize{\textit{Notes:} This index indicates the share of spillovers across countries excluding spillovers to variables within a given country (summed across variable types). Estimated based on rolling windows of varying length. The thicker line is the posterior median alongside the 68 percent posterior credible set (thin lines).}}
    \label{DYindex_RW}
\end{figure}

\begin{figure}[t]
    \includegraphics[width=\textwidth]{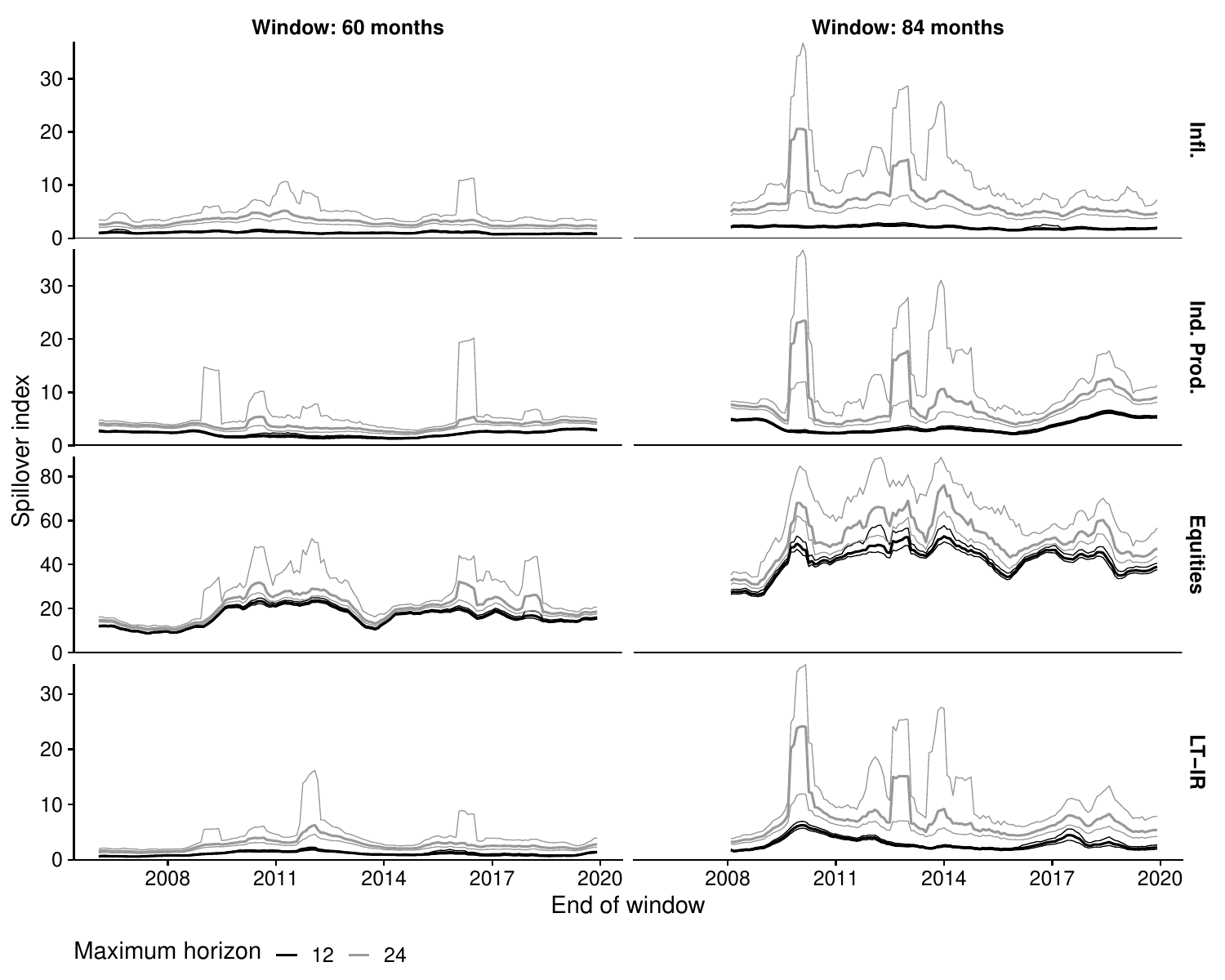}\\ 
    \caption{Diebold-Yilmaz spillover index by variable based on decomposition of forecast error variance of the full system using a rolling window of observations.}\vspace*{-1em}
    \caption*{\footnotesize{\textit{Notes:} This index indicates the share of spillovers between countries one variable at a time (summed across countries). Estimated based on rolling windows of varying length. The thicker line is the posterior median alongside the 68 percent posterior credible set (thin lines).}}
     \label{fig:DYindex_vars_RW}
\end{figure}

\begin{figure}[t]
    \includegraphics[width=\textwidth]{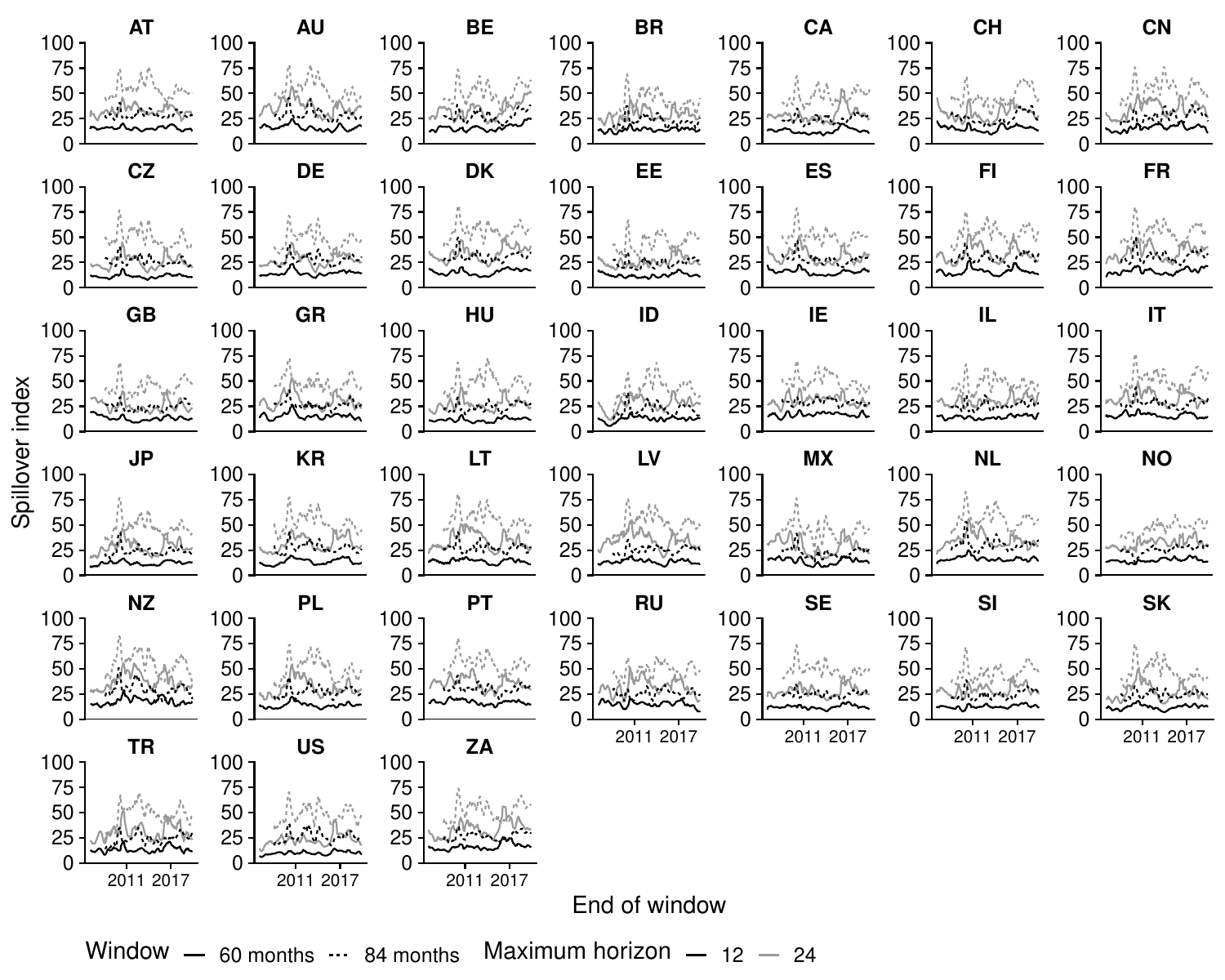}\\ 
    \caption{Diebold-Yilmaz spillover index by country based on decomposition of forecast error variance of the full system using a rolling window of observations.}\vspace*{-1em}
    \caption*{\footnotesize{\textit{Notes:} This index indicates the share of foreign spillovers to the indicated country (summed across variables). Estimated based on rolling windows of varying length. The lines are the posterior median estimates.}}
     \label{DYindex_cts_RW}
\end{figure}

Figures \ref{DYindex_RW} to \ref{DYindex_cts_RW} show the DY indices estimated using rolling windows of varying length rather than an expanding window. In \autoref{DYindex_RW} we show the results of the overall DY index for different rolling window sizes (60 months and 84 months) and for two horizons (12 and 24 months). Two facts that are in line with the literature on using the DY emerge: First, we see a significant increase in spillovers during crisis periods. More specifically, the DY increases strongly during the period from 2009 and 2011. This period was characterized by the emergence of the global financial crisis and -- for Europe -- the sovereign debt crisis. A second peak emerges in 2013/2014, the period of the so-called taper tantrum, i.e., the fear that the U.S. Federal Reserve Bank could wind down its asset purchase program, causing large swings in exchange rates and capital outflows from emerging economies. The third peak occurred around 2016/2017, the period when the Brexit referendum caused severe strains in the European Union and Donald Trump was elected president of the U.S. Second, we see that, in general, spillover indices are larger for higher forecast horizons, which is a finding that is consistent with the analysis provided in Section \ref{sec: emp_work}.

In Figure \ref{fig:DYindex_vars_RW} we show variable-specific spillover indices. We see that for example, spillovers are much larger for equity prices compared to spillovers of inflation, output and long-term interest rates. This finding is also in line with results provided in Section \ref{sec: emp_work}. Spillovers in equity prices increased strongly in the aftermath of the global financial crisis as well as from 2015 onward. In both periods, major central banks launched large scale asset purchase programs which affected financial markets \citep[see e.g.,][]{DAMICO2013425, Rogers2014}. Consistent with our analysis from Section \ref{sec: emp_work}, we also find a peak in spillovers between long-term rates in 2012, which might be explained by the events related to the sovereign debt crisis in Europe. For completeness, we also show country specific spillover indices in \autoref{DYindex_cts_RW}.

\end{appendix}

\clearpage\addcontentsline{toc}{section}{References}
\bibliographystyle{cit_econometrica.bst}
\bibliography{lit}\normalsize\clearpage

\end{document}